\documentclass[aps,twocolumn,showpacs,eqsecnum]{revtex4}

\usepackage{graphicx}
\usepackage{amsfonts}
\usepackage[figuresright]{rotating}  
\usepackage{amssymb}
\usepackage{amsmath}
\usepackage{subfigure}
\usepackage{multirow}
\usepackage{tabularx}

\def\YBCO{YBa$_2$Cu$_3$O$_{6+y}$}

\def\C60{A$_x$C$_{60}$}

\def\HgCu3{HgCa$_2$Cu$_3$O$_{8+y}$}
\def\HgCu4{HgBa$_2$Ca$_3$Cu$_4$O$_{10+y}$}
\def\TlCu{Tl$_2$Ba$_2$CuO$_{6+\delta}$}
\def\TlCu3{Tl$_2$Ba$_2$Ca$_2$Cu$_3$O$_{10+y}$}
\def\TlCu4{Tl$_2$Ba$_2$Ca$_3$Cu$_4$O$_{12+y}$}

\def\BiCu3{Bi$_2$Sr$_2$Ca$_{2}$Cu$_3$O$_y$}

\def\8LSCO{La$_{1.88}$Sr$_{.12}$CuO$_4$}
\def\110LNSCO{La$_{1.5}$Nd$_{0.4}$Sr$_{0.1}$CuO$_{4}$}
\def\stage4LCO{La$_{2}$CuO$_{4+\delta}$}
\def\Y248{YBa$_2$Cu$_4$O$_8$}

\def\NbSe2{NbSe$_2$}
\def\TaSe2{TaSe$_2$}
\def\TiSe2{TiSe$_2$}
\def\NaCoOH2O{Na$_{0.3}$CoO$_{2y}$H$_2$O}
\def\MgB2{MgB${}_2$}

\def\singleRu{Sr$_2$RuO$_4$}

\def\avg#1{\langle#1\rangle}

\def\nn{\nonumber}

\def\me#1#2#3{\langle #1 \vert #2 \vert #3 \rangle}

\begin{document}
\title{Time-Reversal Symmetry Breaking and Spontaneous Anomalous Hall Effect in Fermi Fluids}
\author{Kai Sun}
\affiliation{Department of Physics, University of Illinois at Urbana-Champaign, 1110 West Green Street, Urbana, Illinois 61801-3080, USA}
\author{Eduardo Fradkin}
\affiliation{Department of Physics, University of Illinois at Urbana-Champaign, 1110 West Green Street, Urbana, Illinois 61801-3080, USA}

\begin{abstract}
We study the spontaneous non-magnetic time-reversal symmetry breaking in 
a two-dimensional Fermi liquid without breaking either the translation symmetry or the $U(1)$ charge
symmetry. Assuming that the low-energy physics is described by fermionic
quasiparticle excitations, we identified an ``emergent'' local $U(1)^N$
symmetry in momentum space for an $N$-band model. For a large class of models, 
including all one-band and two-band models, we found that the time-reversal and chiral
symmetry breaking can be described by the $U(1)^N$ gauge theory associated
with this emergent local $U(1)^N$ symmetry. This conclusion enables the classification
of the time-reversal symmetry-breaking states as types $I$ and $II$,
depending on the type of accompanying spatial symmetry breaking. The properties of each class 
are studied. In particular, we show that the states breaking both time-reversal and 
chiral symmetries are described by spontaneously generated Berry phases. We also show 
examples of the time-reversal symmetry-breaking phases in several different microscopically 
motivated models and calculate their associated Hall 
conductance within a mean-field approximation. The fermionic nematic phase with
time-reversal symmetry breaking is also presented and the possible realizations in strongly correlated models such as
the Emery model are discussed.
\end{abstract}
\date{\today}
\maketitle 

\section{introduction}

In this paper we consider the effects of the spontaneous breaking of time-reversal ($\mathbf{T}$) invariance in electronic systems. 
This is  a problem of
considerable current interest particularly in the context of strongly correlated systems.  
While the physics of strong correlation is
important, many aspects of spontaneous time-reversal symmetry breaking are not well understood even at the level of 
weakly coupled systems, well described by Fermi-liquid theory. The problem that we will consider 
is that of the possible quantum phase transitions to states in which time-reversal invariance is spontaneously broken 
in electronic systems with several Fermi surfaces, and how to classify them. 

One of the most important consequences of the spontaneous breaking of time reversal is that these ground states may exhibit
a spontaneous (nonquantized) anomalous Hall effect. More specifically, we consider Fermi systems with multiple Fermi surfaces
with condensates in the particle-hole channel that break time-reversal invariance. As a consequence, these systems have a nontrivial
{\em relative} Fermi-surface Berry curvature which quantifies the strength of the time-reversal symmetry breaking.
The theory that we present here has a close connection with
Haldane's analysis of the anomalous Hall effect as a Berry curvature on the Fermi surface. \cite{Haldane2004}
The nontrivial new effect that results from these states is that they exhibit a spontaneous anomalous Hall effect, 
{\it i.e.\/}, present even in the absence of extrinsic effects such as magnetic impurities or external magnetic fields. 

Time-reversal symmetry breaking in the absence of external
magnetic fields or spontaneous spin ordering has been a focus of interest in condensed-matter physics for quite some time, 
at least since the discovery of high $T_c$ superconductivity in the copper oxide materials. Quite early on it was postulated 
that frustrated two-dimensional (2D) quantum antiferromagnets may have ``chiral spin liquid'' phases (or ground states), 
translationally invariant states without magnetic long-range order.\cite{kalmeyer-1987,Wen1989,Laughlin1988,Fetter1989} 
The known behavior of high $T_c$ superconductors does not appear to be consistent with a spin liquid ground state. The discovery of
time-reversal symmetry-breaking effects in recent experiments on
{\singleRu} and in underdoped {\YBCO} (and similar systems) has renewed the interest in understanding time-reversal 
symmetry-breaking phases in strongly correlated
electronic systems.

The main purpose of this paper is to inquire if it is in  principle possible to 
have an electronic system with a nonmagnetic translationally invariant ground state that 
breaks spontaneously 
time-reversal invariance. Such a system would exhibit a spontaneous Hall effect and 
a Kerr effect even in the absence of disorder and/or magnetic fields.  
We will focus here on simple models of quantum phase transitions in electronic systems, without ferromagnetism 
or any other form of long-range magnetic order. 
We will further assume that these systems 
have well-defined electronic quasiparticle excitations and are hence extensions of Fermi-liquid theory. 
This assumption is valid for any mean-field 
approaches and as will be shown in Sec. \ref{sec:fluctuations}, is 
valid even if the fluctuations of the order parameter around its mean-field 
value are considered for lattice models.
We will see that, even in this ``weak-coupling'' approach, states with the 
desired properties are physically sensible. (Naturally, the naive applicability  of 
the details of this theory to a regime of strong 
correlations, necessary in the context of  the cuprates, is questionable.)

Within this weak-coupling approach, time-reversal breaking phases can be described in terms 
of properties of the resulting one-particle states and of their effective Fermi surfaces. 
Of particular importance is the fact that a system with $N$ Fermi surfaces obtains 
an ``emergent'' (and approximate) local {\em gauge} (in momentum space)
$U(1) \otimes \ldots \otimes U(1)\equiv U(1)^N$ symmetry near the Fermi-liquid fixed point.

In Sec. \ref{sec:gauge} we show that the natural way to represent this structure is in terms of
an Abelian gauge theory of $N$ gauge fields.
These gauge fields describe the quasiparticle Berry phases \cite{Berry1984}
in the sense discussed recently by Haldane. \cite{Haldane2004}
For systems with one or two band(s),  the structure of $\mathbf{T}$ symmetry breaking is
described by these gauge fields and the symmetry properties of the quasiparticle 
dispersion relation under space inversion. However, for more than two bands, more complex $\mathbf{T}$ 
symmetry-breaking phases arise involving additional time-reversal breaking operators which are neither 
Berry connections nor the space inversion symmetry of the quasiparticle dispersion relations. Nevertheless, the 
Berry connections and the inversion symmetry of the quasiparticle dispersion relation still describe a 
large class of the $\mathbf{T}$ symmetry-breaking states even in multiband models. 

We will only consider the $\mathbf{T}$ symmetry-breaking phases 
described by these gauge fields (Berry phases) and the inversion symmetry of quasiparticle dispersion relations.
Within this constraint, the systems that we describe are invariant under the combined transformation of 
$\mathbf{C I T}$, where $\mathbf{C}$ is the chiral transformation ({\it i.e.\/}, a mirror reflection) and
$\mathbf{I}$ stands for space inversion. With the
$\mathbf{C I T}$ symmetry, the $\mathbf{T}$ symmetry-breaking states can be 
classified into two classes, according to the accompanying $\mathbf{C}$ or $\mathbf{I}$ 
symmetry breaking. We refer to the states that preserve all three symmetries $\mathbf{C}$,
$\mathbf{I}$, and $\mathbf{T}$, as the normal states. As for
the $\mathbf{T}$ symmetry-breaking states, if the inversion $\mathbf{I}$ symmetry 
is also broken but 
the chiral $\mathbf{C}$ and the combined $\mathbf{I T}$ symmetries are preserved, 
these states will be referred to as type $I$.
In contrast, the states that break $\mathbf{T}$ and 
$\mathbf{C}$ but 
preserve $\mathbf{C T}$ and $\mathbf{I}$ will be referred as the type $II$ states. The 
states that break all three of $\mathbf{C}$, $\mathbf{I}$, and $\mathbf{T}$ are 
considered as a mixing of types $I$ and $II$. Obviously, type $I$ states have no 
Kerr or Hall effect, but type $II$ may have. The type $I$ state is somehow
trivial if we notice that the momentum $\mathbf{k}$ changes sign under $\mathbf{I}$
or $\mathbf{T}$. Hence, in this paper, we mostly concentrate on the type $II$ states.

Using the Berry connections, we show that type $I$ phases may appear in one-band 
or multiband 
models. However, the type $II$ phases can only be found in multiband models.
The phase transitions from the normal phase to the type $II$ phase can be classified 
into two different scenarios depending on whether the band structure has degeneracy 
points or not (degeneracy lines or areas usually require fine tuning and will not be 
considered).

After exploring the general theory, we use mean-field theory to investigate the 
$\mathbf{T}$ symmetry
breaking in several specific models.
In Sec. \ref{sec:order_parameter}, we study the general symmetry properties 
of the (fermion bilinear) order parameters in 
the particle-hole channel for systems whose band structure contains no 
degeneracy point and show that the type $II$ $\mathbf{T}$ symmetry-breaking 
states requires two order parameters.

In Sec. \ref{sec:isotropic}, we present a mean-field study of
2D Fermi liquids with continuous rotational and translational
symmetries, and time-reversal invariance. In this section we discuss the possible patterns of spontaneous breaking of
time-reversal invariance, inversion and chiral symmetries, and rotational invariance in interacting metallic systems. 
Although the models we discuss here use the framework of the Landau theory of the Fermi liquid, 
the patterns of symmetry breaking
that are found, as well as the resulting phenomenology, are of more general interest.
In a Landau-type model with four-fermion forward
scattering interactions, analogous to the type discussed 
in Refs. \cite{Oganesyan2001} and \cite{Wu2007},
time-reversal symmetry-breaking phases
can be stabilized and are usually accompanied by rotational symmetry breaking. 
In a one-band model, 
the type $I$ phases can be reached through a Pomeranchuk 
instability \cite{Pomeranchuk1959} 
in {\em odd angular momentum channels} (with angular momentum $\ell >1$), 
nematic-like phases with broken space inversion and time reversal. 
However for two-band models type $II$ 
phases may also appear, and  have a similar structure to the  $\beta$ phases   
in fermionic systems with spin described in Refs.[\onlinecite{Wu2004}] 
and [\onlinecite{Wu2007}]. 
In this  section we construct the phase diagram. Here we also evaluate the Hall conductance 
for these phases, which is not quantized since these phases 
are gapless and conducting. We also show that the Hall conductance found here is related with a 
topological index ,the Kronecker index of the homotopy mappings $S^1\rightarrow S^1$, {\it i.e.\/}, $\pi_1(S^1)$.
This in turn implies that the $\mathbf{T}$ symmetry breaking in these phases
is stable against adiabatic perturbations, even though the actual value of the 
{\em unquantized} Hall conductance is not universal and can be changed continuously.

In Sec. \ref{sec:lattice_model}, we generalize these $\mathbf{T}$ symmetry-breaking phases to
lattice models and discuss subtle effects arising from the degeneracy points in the band structures.
In particular we show that without degeneracy points, 
the time reversal $\mathbf{T}$ and chiral $\mathbf{C}$ symmetry-breaking phase can be reached from 
a normal Fermi liquid either by a direct first-order transition or by two separate phase transitions 
through an intermediate phase characterized by rotational symmetry breaking. But in the presence of degeneracy points,
the direct transition between the $\mathbf{T}$ and $\mathbf{C}$ symmetry-breaking phase and the normal Fermi liquid may 
be second order.

Finally, in Sec. \ref{sec:discussion}, we present a discussion of the experimental consequences of this work.
The relation  between this work and 
its particle-particle channel counterpart is also discussed, as well as  the similarities 
and differences  with the phases studied in Ref. \cite{Wu2007}. 
We present details of the calculations for two-band models in Appendix
\ref{app:two_band}. The topological and physical meaning of the Wilson loops introduced 
in Sec. \ref{sec:gauge} is discussed in
Appendix \ref{app:wilson_loop}. In Appendix \ref{app:hall_cond}, we present the 
details of the calculation of the Hall conductivity, and in Appendix \ref{app:Z2}, the symmetry
of the $\alpha_2$ and $\beta_2$ phases (to be defined below) is discussed.

\section{Gauge theory and Berry phase}
\label{sec:gauge}

In this section, we study the general properties of the spontaneous $\mathbf{T}$ symmetry 
breaking for a fermionic system, which we will assume to be well described by an effective Fermi
liquid, {\it i.e.\/}, a fermionic system with well-defined quasiparticle excitations which are asymptotically free at low
energies. We will also assume that time-reversal invariance is not broken explicitly and hence
that there is neither an external magnetic field nor any sort of magnetic long-range order. 
We will consider systems without magnetic impurities, trapped magnetic fluxes, or other explicit extraneous time-reversal symmetry-breaking effects. 
For one- and two-band models, we will show that:
\begin{enumerate}
	\item 
		The $\mathbf{T}$ symmetry-breaking effects are represented either by the existence of Berry phases or by the 
		symmetry properties of the quasiparticle dispersion relation under space inversion;
	\item 
		There is a $\mathbf{CIT}$ symmetry, where $\mathbf{T}$ is time-reversal,
		$\mathbf{C}$ is a chiral transformations (reflection across a suitable mirror plane),
		and $\mathbf{I}$  is space inversion;
	\item 
		In the absence of explicit breaking of $\mathbf{T}$, 
		the total Berry phase of all the bands is zero, 
		$\sum_n \Phi_{\Gamma}^n=0$;
	\item 
		There is no type $II$ $\mathbf{T}$ symmetry breaking in a one-band model;
	\item 
		Degeneracy points of the effective band structure (defined later) 
		have an associated quantized Berry flux $n\pi$, with integer $n$;
	\item 
		Systems with and without degeneracy points have different properties
                when undergoing a phase transition to a type $II$ $\mathbf{T}$ 
		symmetry-breaking phase.
\end{enumerate}
Here type $I$ (breaking $\mathbf{I}$ and $\mathbf{T}$) and type $II$ (breaking $\mathbf{C}$ and $\mathbf{T}$) refer to the two types of $\mathbf{T}$ breaking phases discussed in the Introduction.
For models with more than two bands, we will show that by assuming $1$, all other 
conclusions above can be generalized easily.

The low-energy properties of a Fermi liquid are described by its spectrum of quasiparticle 
excitations, {\it i.e.\/}, Bloch waves and their dispersion relation. The dispersion relation 
$\epsilon_n(\mathbf{k})$, where $n$ is the band index with $n=1,2,\ldots,N$ for an $N$ band model,
transforms to $\epsilon_n(-\mathbf{k})$ under time reversal $\mathbf{T}$ or space inversion $\mathbf{I}$
but is invariant under chirality $\mathbf{C}$ or the simultaneous action of $\mathbf{T}$ and $\mathbf{I}$ .
Hence, the odd part of the dispersion relation, $\epsilon_n(\mathbf{k})-\epsilon_n(-\mathbf{k})$, describes 
type $I$ $\mathbf{T}$ symmetry breaking.
 
The Bloch waves may also contain information of the $\mathbf{T}$ symmetry breaking.
Due to the (perturbative) irrelevance at low energies of the quasiparticle interactions 
under the renormalization
group (RG), \cite{Shankar1994} a Fermi liquid with $N$ bands is invariant under a 
$U(1)^N$ gauge transformation
\begin{align}
|\psi_n(\mathbf{k})\rangle\rightarrow e^{i\varphi_n(\mathbf{k})}|\psi_n(\mathbf{k})\rangle,
\label{eq:gauge-transf1}
\end{align}
where $|\psi_n(\mathbf{k})\rangle$ is the Bloch wave function of the band $n$. 
This $U(1)^N$ gauge symmetry, associated with independent redefinitions (gauge
transformations) of the phase of the quasiparticle Bloch states for each band at each wave vector
$\mathbf{k}$,  is
an ``emergent'' symmetry, asymptotically accurate only close enough to the Fermi-liquid fixed point. Away from this
fixed point, the irrelevant quasiparticle scattering processes 
make this $U(1)^N$ symmetry an approximate one. This effect can be studied perturbatively as will be 
shown in Sec. \ref{sub:sec:other_scatterings}, but can (and will) 
be ignored for the purposes of the present discussion.

To remove the redundant degrees of freedom, one defines the overlap matrix \cite{Blount1962}
\begin{align}
\mathcal{A}^a_{nm} = -i \langle\psi_n(\mathbf{k})|\nabla_\mathbf{k}^a \psi_m(\mathbf{k})\rangle.
\end{align} 
The diagonal terms, $\mathcal{A}_{nn}^a$, are Berry connections which under the gauge
transformation of Eq.\eqref{eq:gauge-transf1} transform as gauge fields:
\begin{align}
\mathcal{A}^a_{nn} \rightarrow \mathcal{A}^a_{nn}+\nabla^a_{\mathbf{k}}\varphi_n,
\label{eq:gauge-transf2}
\end{align}
The off-diagonal terms, for $n \neq m$, transform instead as
\begin{align}
\mathcal{A}^a_{nm}\rightarrow e^{-i\varphi_n}\mathcal{A}^a_{nm}e^{i\varphi_m},
\label{eq:gauge-transf3}
\end{align}
which cannot be regarded as gauge fields.
Clearly the overlap matrix $\mathcal{A}^a_{nm}$ are the matrix elements of 
the position operator in Bloch states. \cite{Blount1962} The diagonal terms
are directly related with the ``anomalous velocity'' in the semiclassical theories of 
Bloch waves. \cite{Sundaram1999}

It is well known from the theory of the Hall effect 
(see, for instance, Haldane's work\cite{Haldane2004} and references
therein) that an external magnetic field induces a nontrivial Berry curvature. 
This affects all the bands in essentially the same way. In what follows we will assume that time-reversal invariance is not broken explicitly by external fields. Thus, the total Berry curvature vanishes
as required by Eq. \eqref{eq:constrain}.

We now restrict our discussion on $\mathbf{T}$ symmetry breaking that can be described 
by the diagonal terms $\mathcal{A}_{nn}$ alone.  As shown in Appendix \ref{app:two_band}, 
this assumption is automatically satisfied for all one- and two-band models. For systems with three or more bands 
it is also possible to have phases that break time-reversal invariance which are described purely by off-diagonal
operators, $\mathcal{A}_{nm}$ (with $n \neq m$). Even though they do break time-reversal, these states do not have a Berry phase and, 
consequently, will not have a spontaneous anomalous  Hall effect. We will discuss in Sec.
\ref{sec:lattice_model} that one of the so-called Varma loop states, $\theta_I$ in the notation of
Ref.[\onlinecite{Varma2006}], is an example of a time-reversal symmetry-breaking state in a three-band model without an anomalous Hall effect.

From now on, only the diagonal terms $\mathcal{A}_{nn}$ will be considered. Their,
gauge-invariant, physical degrees of freedom are described by the Wilson loops
\begin{align}
W^n_{\Gamma}=\exp (i \Phi^n_{\Gamma}),\\
\Phi^n_{\Gamma}=\oint_{\Gamma} \sum_a \mathcal{A}^a_{nn} d k^a,
\end{align}
which is also the Berry phase. \cite{Haldane2004} Notice that in principle one can consider any path
$\Gamma$ in momentum space. In practice, for a gapless system the path of physical interest  coincides
with the location of the Fermi surfaces of the bands (see below).

Under the action of the chiral transformation $\mathbf{C}$, space inversion $\mathbf{I}$, and 
time reversal $\mathbf{T}$, the Wilson loops $W^n_{\Gamma}$ transform as
\begin{align}
\mathbf{C} W^n_{\Gamma}=(W^n_{\Gamma})^*,\\
\mathbf{I} W^n_{\Gamma}=W^n_{\mathbf{I} {\Gamma}},\\
\mathbf{T} W^n_{\Gamma}=(W^n_{\mathbf{I} {\Gamma}})^*.
\end{align}
The chiral transformation $\mathbf{C}$ reverses the orientation of the path, ${\Gamma}\rightarrow -{\Gamma}$,
and hence, it transforms $W^n_{\Gamma}$ into its complex conjugate.
The space inversion operator $\mathbf{I}$ changes momentum $\mathbf{k}$ to  $-\mathbf{k}$, which changes
the integration contour but preserves its direction. The time-reversal operator $\mathbf{T}$ is antiunitary.
It changes $\mathbf{k}$ to  $-\mathbf{k}$ as $\mathbf{I}$ but also changes the Wilson loop to its complex conjugate, which is also how  $\mathbf{C}$ acts. Therefore, $W^n_{\Gamma}$ is invariant under $\mathbf{C I T}$. Since the
dispersion relation is invariant under $\mathbf{I T}$ (and $\mathbf{C}$), then the system
must be invariant under $\mathbf{C I T}$. This is one
of the key results of this paper, which makes the classification
of the types $I$ and $II$ $\mathbf{T}$ symmetry-breaking states possible.

As we noted above, in the absence of magnetic fields (and of any other explicit breaking of time
reversal), there is a constraint over the total Berry phase,
\begin{align}
	\sum_n \Phi^n_{\Gamma}=0,
	\label{eq:constraint2}
\end{align}
which implies that there is 
no Berry phase associated with the charge sector, the overall $U(1)$ gauge group. 
However, in the presence of magnetic fields, the phase of the Bloch wave cannot be
determined in a unique and smooth way over the entire Brillouin zone. \cite{Kohmoto1985} This  
invalidates the assumptions behind Eq. \eqref{eq:constrain}, leading to
a nontrivial Berry phase in the charge $U(1)$ sector and
a nonvanishing Hall conductance.
On the other hand, in the absence of external magnetic fields, although the constraint of Eq. 
\eqref{eq:constraint2} prevents the charge $U(1)$ sector to obtain a Berry phase,
a nontrivial {\em relative Berry phase} between different bands is still allowed. This 
is the key point in our study of a spontaneous $\mathbf{T}$ symmetry breaking without
magnetic ordering that we are interested in here.

For a one-band model, the constraint that the total Berry phase must be trivial,  
$\sum_n \Phi^n_{\Gamma}=0$, implies that the Wilson loops must  be real, 
$W_{\Gamma}=W_{\Gamma}^*$. Hence, the Wilson loop is an eigenvector of the chiral transformation $\mathbf{C}$, 
{\it i.e.\/}, 
$\mathbf{C} W_\Gamma=W_\Gamma$, and  the $\mathbf{I T}$ symmetry always holds. Thus, for one-band
models, only type $I$ time-reversal spontaneous symmetry breaking
is allowed.

Let us consider now the case of two-band models which allow for a richer structure. 
In this case Eq.\eqref{eq:constraint2} now becomes
$\Phi^1_{\Gamma}+\Phi^2_{\Gamma}=0$. Hence, only one Berry phase $\Phi^1_{\Gamma}$ 
(or equivalently $\Phi^2_{\Gamma}$) is linearly independent. 
For a state with $\mathbf{C}$, or equivalently $\mathbf{I T}$, symmetry (the normal state
or the type $I$ state), again one obtains $W^1_{\Gamma}=(W^1_{\Gamma})^*$, 
which quantizes the Berry phase to 
be $\Phi^1_{\Gamma}= n \pi$, with integer $n$. Conversely,  if $\mathbf{C}$ {\em and} $\mathbf{I T}$ 
are broken, the symmetry breaking is type $II$. In this case,  $\mathbf{T}$ symmetry breaking in two-band models
can be described by $W^1_{\Gamma}$ obtaining an imaginary part. Equivalently, the Berry phase  
$\Phi^1_{\Gamma}$ becomes 
nonquantized for a type $II$ time-reversal symmetry breaking. 

The effective one-particle Hamiltonians of a two-band model has the form of a $2 \times 2$ Hermitian 
matrix whose coefficients are smooth functions of the momentum $\mathbf{k}$. In Appendix \ref{app:wilson_loop} 
we use standard arguments\cite{Berry1984,Avron1983} to relate the Berry phase for this system to a Wess-Zumino 
term, familiar from the path integral for spin (see, {\it e.g.\/}, Ref. [\onlinecite{Fradkin1991}]). 
The Berry phase actually equals half of the Wess-Zumino term. For a specific contour, where $\Gamma$ 
coincides with the Fermi surface, the
Berry phase is proportional to the Hall conductance, \cite{Haldane2004}
as reviewed in Appendix \ref{app:hall_cond}. For an insulator, which does not have a Fermi surface, 
the contour is the boundary of the Brillouin zone. This leads to the well-known quantization of 
the Wess-Zumino term as a Chern number, and it implies the quantization of the Hall conductance.
However, for a metallic state, the contour $\Gamma$ is the Fermi surface, and the manifold is no longer 
compact. As emphasized by Haldane,\cite{Haldane2004} in this case the Hall conductance is in general not 
quantized, which leads to anomalous Hall effect. In Sec. \ref{sub:sec:hall_cond},
the anomalous Hall conductance for the metallic spontaneous type II $\mathbf{T}$ symmetry-breaking phase
in a specific model will be computed within a mean-field approximation.

We end the discussion of this section by emphasizing one property of the Berry phase for later use.
For systems with $\mathbf{I T}$ symmetry, {\it i.e.\/}, $\Phi^n_{\Gamma}= n \pi$, field strength 
\begin{align}
	\mathcal{F}^{ab}_n=\nabla_{\mathbf{k}}^a \mathcal{A}^b_n-\nabla_{\mathbf{k}}^b \mathcal{A}^a_n,
\end{align}
is zero away from degeneracy points ({\it i.e.\/}, for points 
in momentum space where the bands have different energy). However, at degeneracy points ({\it i.e.\/}, points 
in momentum space where two or more bands have the same energy) the field strength may have Berry flux
$n \pi$, with integer $n$. As will be shown later, this difference of the Berry phase leads
to different phase transitions in type $II$ $\mathbf{T}$ symmetry breaking.

In the rest part of this paper, we will study specific microscopic models where the conclusions above are applied. 

\section{order-parameter theory without degeneracy point}
\label{sec:order_parameter}

In this section, we study systems without degeneracy point in the band structure by writing down 
order parameters in the particle-hole channel that preserve both translational and charge $U(1)$ symmetries,
the most general ground-state expectation values of bilinears in fermion 
operators of the form
\begin{equation}
O=\sum_{\mathbf{k},n,m} 
\me{\textrm{gnd}}{\psi^{\dagger}_{n}(\mathbf{k})M^{nm}(\mathbf{k})\psi_{m}(\mathbf{k})}{\textrm{gnd}},
\label{eq:fermion-bilinears}
\end{equation}
where $\psi_n^\dagger(\mathbf{k})$ and $\psi_n(\mathbf{k})$ are the fermionic creation and annihilation 
operators and $\mathbf{k}$ is the momentum of the quasiparticle; the indices $n$ and $m$  label the bands. 
The spin indices are dropped since we are not considering spin ordering. Each Hermitian matrix 
$M$ defines a real order parameter $O$
(complex order parameters are two real order parameters).

In the absence of band crossings, the $\mathbf{I T}$ symmetry implies the 
existence of a special gauge in which $\mathcal{A}^a_{nn}=0$. With this gauge choice, under 
$\mathbf{I}$ and $\mathbf{T}$,
$M(\mathbf{k})$ transforms as
\begin{align}
\mathbf{I} M(\mathbf{k}) \mathbf{I}^{-1}&=M(-\mathbf{k}),
\\
\mathbf{T} M(\mathbf{k}) \mathbf{T}^{-1}&=M^*(-\mathbf{k}).
\end{align}

The fermion bilinears defined in Eq.\eqref{eq:fermion-bilinears} can always be expressed in such a 
way that transform irreducibly under the symmetries of the system. 
Here we are interested in particular in their transformation properties under time reversal.
In mean-field theories, such as the one we will discuss in Sec. \ref{sec:isotropic}, the one-particle
effective Hamiltonian depends linearly on these fermionic bilinear order-parameter fields. Thus, a
non-vanishing expectation value of the order parameter breaks the symmetry. 

It is not possible to have a state that breaks spontaneously time-reversal invariance in a way that cannot
be compensated by another symmetry transformation. In Sec. \ref{sec:single} we will construct a state
in one-band model with a ground state that breaks $\mathbf{T}$ which however must also break $\mathbf{I}$,
space inversion. This model, which has an order parameter in the particle-hole channel 
with angular momentum $\ell=3$, is an explicit representation of the ``Varma loop'' state ($\theta_{II}$) discussed recently by Varma as a $\mathbf{T}$ breaking state\cite{Varma2006}. 
The state is, however, also odd under $\mathbf{I}$ but invariant under $\mathbf{IT}$ and under chirality $\mathbf{C}$. 
Thus, this is a type $I$ state in the classification discussed in the Introduction. As a consequence this state
does not have a spontaneous anomalous Hall effect or a uniform Kerr effect in the absence of external
magnetic fields or defects.
(In this sense this pattern of time-reversal symmetry breaking is analogous to that of a N\'eel
antiferromagnet, in which time reversal and translation invariance by one lattice constant are 
broken but the combination of both is not.)

In order to obtain a state with time-reversal symmetry breaking but without space inversion symmetry breaking, 
it is necessary to have at least 
two bands. For two-band models, even though the natural symmetry in this case is $U(1) \times U(1)$, 
it can be naturally embedded in a the larger $U(2)$ group (even though it is not a symmetry). To obtain a
ground state with broken time-reversal invariance it is necessary to break  $U(2)$ completely 
(down to its center, the $\mathbb{Z}_2$ subgroup). This requires that two (non-commuting) generators of 
$U(2)$ must be broken in the ground state. Thus the order parameter has two components and cannot be made
real by a gauge transformation. Similarly, the eigenstates of the effective one-particle Hamiltonian are
complex. A system with these properties will have a nontrivial Berry connection. It turns out that it also breaks
chirality, $\mathbf{C}$, but it is invariant under $\mathbf{CT}$. Hence, this state  corresponds to a type
$II$ time-reversal symmetry breaking. A state of this type has a spontaneous anomalous Hall effect and a
Kerr effect even in the absence of external magnetic fields and defects.
Examples of states of this type in the particle-particle channel are the well-known 
$\mathbf{T}$ breaking spin triplet $p_x+ip_y$ and spin singlet $d_{x^2-y^2}+i d_{xy}$ 
superconducting condensates. Similarly, the $d_{x^2-y^2}+i d_{xy}$ $d$-density wave ($d$DW) state breaks 
time reversal and translation invariance [with $\mathbf{Q}=(\pi,\pi)$] in the particle-hole channel.

In the following sections we present some simple models containing the types $I$ and $II$ 
$\mathbf{T}$ symmetry-breaking states.
For simplicity, only one- and two-band models will be studied. However, the 
conclusion can be generalized to multiband models with little effort.

\section{Rotationally Invariant Models}
\label{sec:isotropic}

Let us consider a 2D isotropic fermionic system with Hamiltonian
\begin{align}
H=\sum_{\mathbf{k},n}
\psi^\dagger_n(\mathbf{k})(\epsilon_n(\mathbf{k})-\mu)
\psi_n(\mathbf{k})
+H_{\textrm{int}},
\end{align}
where $\epsilon_n(\mathbf{k})$ is the single-particle kinetic energy for band
$n$ and $\mu$ is the chemical potential. $H_{\textrm{int}}$ is the 
interacting part of the Hamiltonian. Here we only consider the forward-scattering interactions. 

\subsection{One-band model}
\label{sec:single}

\begin{figure}
\begin{center}
\subfigure[$\;\ell=2$]{\includegraphics[width=0.22\textwidth]{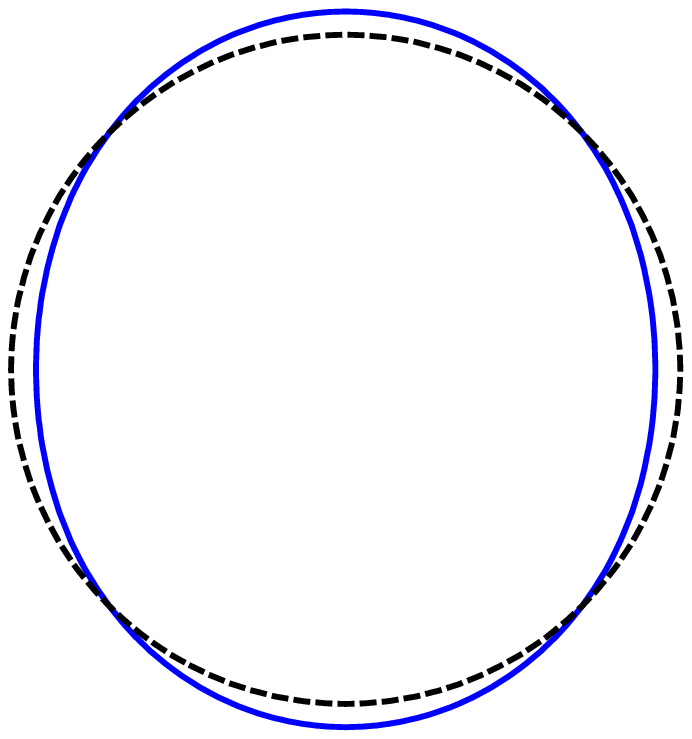}}
\subfigure[$\;\ell=3$]{\includegraphics[width=0.22\textwidth]{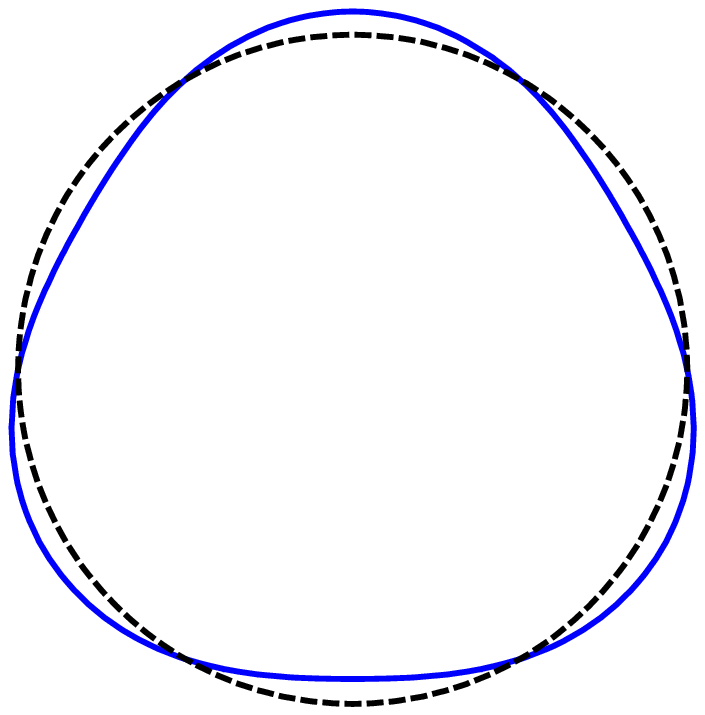}}
\subfigure[$\;\ell=4$]{\includegraphics[width=0.22\textwidth]{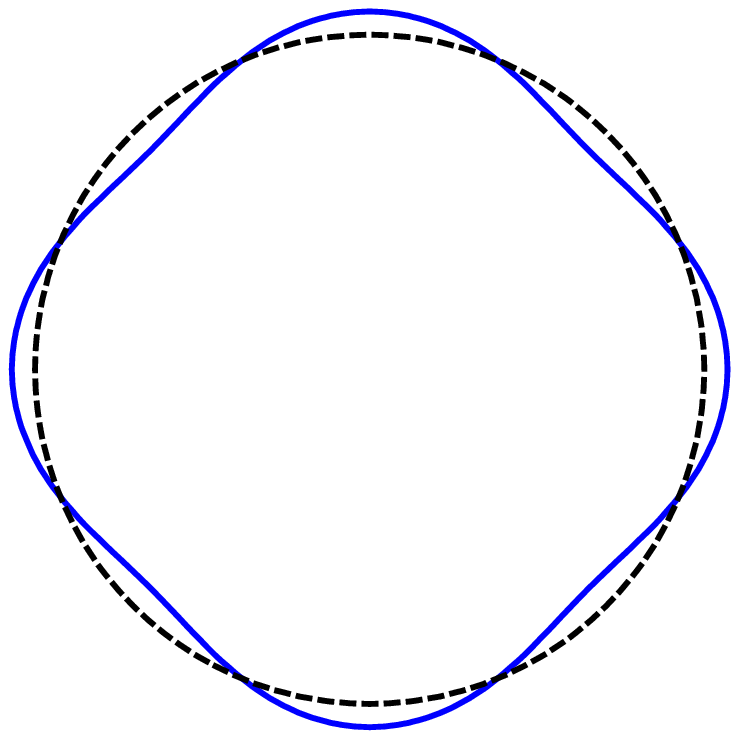}}
\subfigure[$\;\ell=5$]{\includegraphics[width=0.22\textwidth]{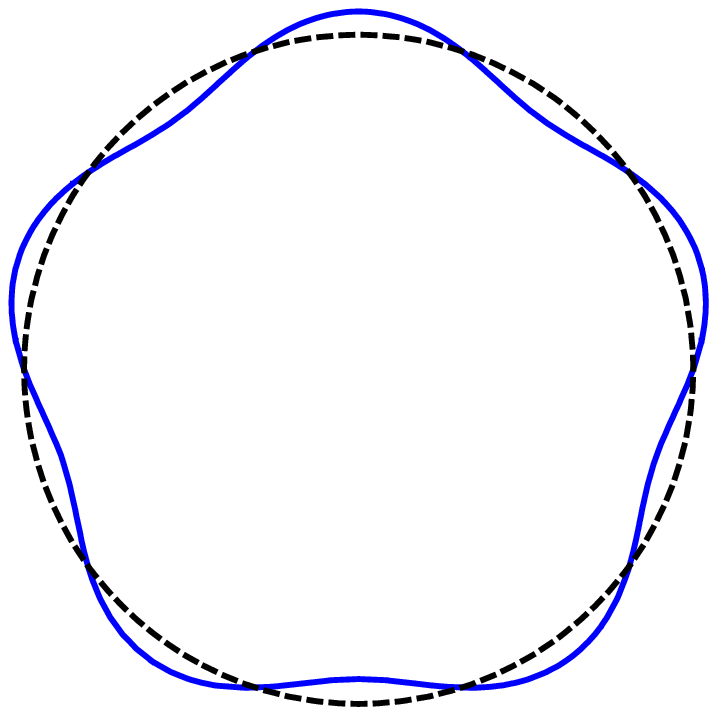}}
\end{center}
\caption{(Color online) Fermi surfaces of phases with Pomeranchuk instability.
Odd $\ell$ states break the $\mathbf{I}$ and $\mathbf{T}$ symmetries but preserve $\mathbf{I T}$.
}\label{fig:pomeranchuk}
\end{figure}

In a one-band model, $H_{\textrm{int}}$ can be expanded into different 
angular momentum channels, denoted by the angular momentum quantum number 
$\ell$. If the coupling constant 
in some channel, $f_\ell$, is attractive and strong enough to violate the Pomeranchuk condition, 
$f_\ell N(0)+2<0$, where $N(0)$ is the density of states at the Fermi 
level, then the spherical Fermi surface becomes unstable, leading to a spontaneously distorted 
Fermi surface as shown in Fig. \ref{fig:pomeranchuk}. This is the 
Pomeranchuk instability. \cite{Pomeranchuk1959}
The case of the $\ell=2$ channel is the electron nematic phase. \cite{Kivelson1998,Oganesyan2001} Similar condensates for
other even angular momentum channels can also be (and have been) considered. 

In the case of  odd angular momentum channels both time reversal $\mathbf{T}$ and 
space inversion $\mathbf{I}$  are spontaneously broken, but the product 
$\mathbf{TI}$ remains unbroken as these states are not chiral: $\mathbf{C}$ is unbroken. 
These states correspond to the type $I$ $\mathbf{T}$ symmetry-breaking states.
As can be seen from Figs. \ref{fig:pomeranchuk}(b) and (d), the chiral
symmetry is preserved and the $\mathbf{T}$ symmetry can be recovered by space
inversion.

\subsection{Two-band model with $U(1)\otimes U(1)$ symmetry}
\label{sub:sec:one_scattering}

Let us now consider the case of a two-band model. It can be either a system with two bands due to band-structure 
effects or a bilayer system with a small amount of hybridization (the bonding and antibonding Fermi surfaces are 
close). We will assume that the system has full translation and rotation symmetries (which will reduce to a 
point-group symmetry for a lattice system). As before (and for simplicity) we will assume that the system is nonmagnetic (so that
the spin degrees of freedom yield only a redundant effect) and that the Fermi surfaces are such that the system is not unstable to the formation of charge density waves or any other instability. We will also ignore possible superconducting
states.

We will also assume that the separation between the two bands is small and only the 
forward-scattering channels need to be considered. Under these approximations, the 
scattering processes can be classified by their angular momentum 
channels, as in the case of the one-band model, and we can define bosonic fields (fermion bilinears)
\begin{align}
\phi_{\ell 1,\mu}(\mathbf{q})=\sum_{\mathbf{k},n,m}:\psi^\dagger_n(\mathbf{k}+\frac{\mathbf{q}}{2}) 
\cos(\ell \theta_k) 
\sigma_\mu^{nm}\psi_{m}(\mathbf{k}-\frac{\mathbf{q}}{2}):,
\\
\phi_{\ell 2,\mu}(\mathbf{q})=\sum_{\mathbf{k},n,m}:\psi^\dagger_n(\mathbf{k}+\frac{\mathbf{q}}{2})
\sin(\ell \theta_k) 
\sigma_\mu^{nm}\psi_{m}(\mathbf{k}-\frac{\mathbf{q}}{2}):.
\end{align}
Here $:\psi^\dagger\psi:$ stands for the normal-order product, relative to the ground state of a Fermi liquid (or gas).
Here $n,m=1,2$ label the two bands. The matrices $\sigma_{\mu}$ are the identity matrix for $\mu=0$ and the 
Pauli matrices for $\mu=x$, $y$, or $z$. $\theta_k$ is the polar angle of the momentum vector $\mathbf{k}$.

With these definitions, $H_{\textrm{int}}$ can be written as the sum of all 
quadratic terms in $\phi_{\ell 1,\mu}$ and $\phi_{\ell 2,\mu}$ that preserve momentum and 
angular momentum. We further assume that the instability only occurs in one 
angular momentum channel $\ell$ so that collective excitations in all other angular momentum 
channels are gapped and irrelevant at low energies. For now, we 
only consider one particular interaction Hamiltonian $H_{\textrm{int}}$ of the form
\begin{align}
H_{\textrm{int}}=\sum_{\mathbf{q}}\frac{f_\ell(q)}{2}
\sum_{i=1,2}\sum_{\mu=x,y}\phi_{\ell i,\mu}(\mathbf{q})\phi_{\ell i,\mu}(-\mathbf{q}).
\label{eq:fs_xy}
\end{align}
Written in terms of fermionic operators, we can see that this interaction corresponds to the scattering channel
$:\psi^\dagger_1\psi_2::\psi^\dagger_2\psi_1:$.
Other scattering channels will be studied later in Sec. \ref{sub:sec:other_scatterings}.

In addition to the $U(1)$ charge symmetry, this Hamiltonian has an extra 
internal $U(1)$ symmetry 
corresponding to the relative phase between the two bands. This is
because the interaction $H_{\textrm{int}}$  preserves particle number 
in each band. This high symmetry requires some amount of fine tuning, but 
as we will show later, most of the properties are preserved even in the absence of 
this symmetry (at least perturbatively).

Just as in the case of  the one-band model discussed in Sec. \ref{sec:single}, if an interaction in some
angular momentum channel is attractive and in magnitude exceeds a critical value, the ground state of the system 
becomes unstable.  The corresponding order parameters can be taken to be two two-component real vectors
in the $U(1)$ relative phase manifold
\begin{align}
\vec{\phi}_{\ell i}=\left(\avg{\phi_{\ell i,x}(\mathbf{q}=0)},\avg{\phi_{\ell i,y}(\mathbf{q}=0)}\right)
\end{align}
with $i=1$ or $2$. Notice that we use bold characters to represent vectors in space (or momentum space)
but use $\vec{\phi}$ to indicate the two two-component real vector order parameters which form a representation of
the nondiagonal piece of $U(1) \otimes U(1)$ group.

\begin{figure}
\begin{center}
\includegraphics[width=0.4\textwidth]{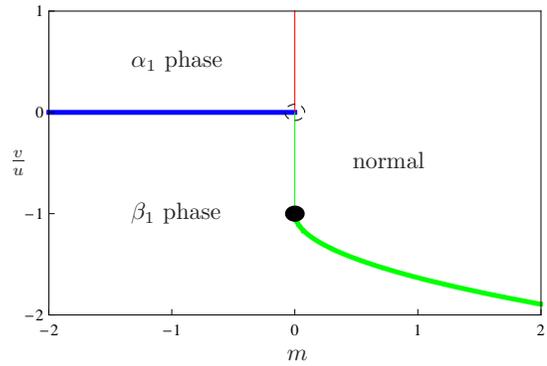}
\end{center}
\caption{(Color online) The phase diagram with exact relative $U(1)$ symmetry.
The red line is the phase boundary between the $\alpha_1$ and the normal phases.
The green ones are between the $\beta_1$ and the normal phases, and the blue
one is between the $\alpha_1$ and the $\beta_1$ phases. Thick lines are first
order phase boundaries and others are second-order ones. The black
dot is a tricritical point. The dashed
circle  is a critical point where the first-order phase
boundary meets two second-order phase transitions.}
\label{fig:phase1}
\end{figure}

In order to preserve the spatial symmetries and the internal $U(1)$ symmetry, 
the Landau free energy has the form
\begin{align}
F=m(|\vec{\phi}_{\ell 1}|^2+|\vec{\phi}_{\ell 2}|^2)+u(|\vec{\phi}_{\ell 1}|^2+|\vec{\phi}_{\ell 2}|^2)^2
\nn\\
+4v(|\vec{\phi}_{\ell 1} \times \vec{\phi}_{\ell 2}|)^2+\textrm{higher order terms},
\end{align}
This free energy is very similar to the spin Pomeranchuk instability 
states in Ref. \cite{Wu2007}, except that the internal symmetry here is 
the relative phase $U(1)$ instead of the spin $SU(2)$. The resulting mean-field
phase diagram for the system at hand is shown in Fig. \ref{fig:phase1}. 

The coefficients of the free energy can be  determined by a mean-field calculation in the same spirit as that of
Ref.\cite{Wu2007}. We obtain
\begin{align}
m=&-\left(\frac{N(0)}{4}+\frac{1}{2 f_\ell(0)}\right)
\nn\\
&+\Delta^2 \frac{N(0)}{96}
\left[3\left(\frac{N^{\prime}(0)}{{N(0)}}\right)^2-\frac{N^{\prime\prime}(0)}{{N(0)}}\right]
\\
u=&\frac{N(0)}{64}
\left[2\left(\frac{N^{\prime}(0)}{{N(0)}}\right)^2-\frac{N^{\prime\prime}(0)}{{N(0)}}\right],
\label{eq:mean_field_u}
\\
v=&\frac{N^{\prime\prime}(0)}{48}.
\label{eq:mean_field_v}
\end{align}
Here $\Delta$ is the energy splitting between the two bands, which is assumed to be much 
smaller than the Fermi energy $\epsilon_F$, $\Delta \ll \epsilon_F$; $N(0)$ is the density of states at the 
Fermi surface calculated using the average dispersion relation $[\epsilon_1(\mathbf{k})+\epsilon_2(\mathbf{k})]/2$. 
$N^{\prime}(0)$ and $N^{\prime\prime}(0)$ are the first- and second-order derivatives of the density of states
$N(\epsilon)$ at the
Fermi surface.
Higher order terms will be needed for stability reasons if $u<0$ or $u+v<0$. For
simplicity, we only consider $u>0$ and assume that the higher order term
is $w (|\vec{\phi}_{\ell 1}|^2+|\vec{\phi}_{\ell 2}|^2)^3$ with $w>0$.

We will now discuss the structure of the phase diagram of Fig.\ref{fig:phase1}. The system has three phases: a) the
normal phase with $\vec \phi_{\ell 1}=\vec \phi_{\ell 2}=0$, b) the $\alpha$ phase in which 
$\vec \phi_{\ell 1} \times \vec \phi_{\ell 2}=0$, and c) 
the $\beta$ phase in which they are orthogonal, $\vec \phi_{\ell 1} \cdot \vec \phi_{\ell 2}=0$ and 
$|\vec \phi_{\ell 1}|=|\vec \phi_{\ell 2}|$.
For $u>0$ and $u+v>0$, $m=0$ 
marks the second-order phase boundary between the normal state and either the $\alpha$ or the $\beta$ phase, 
depending on the sign of $v$. 

To distinguish this $\alpha$ phase with
other similar phases which will be discussed below, we indicate this phase
as $\alpha_1$ in the phase diagram of Fig. \ref{fig:phase1} (same with the $\beta$ phase). 
This phase has a distorted Fermi surface, as shown in Fig. 
\ref{fig:FS_alpha_1}.  In the $\alpha_1$ phase, the Fermi surfaces of both bands are distorted, and their
rotational symmetries are reduced from $SO(2)$ to a $2\ell$-fold discrete  symmetry. 
The $\alpha_1$ phase preserves the $\mathbf{T}$ symmetry
as well as $\mathbf{I}$ and $\mathbf{C}$. A similar phase on the square lattice is 
discussed in Ref. [\onlinecite{Puetter2007}], where it is referred to as the ``hidden nematic phase.''

For $v<0$, the system is in the $\beta_1$ phase. In this phase, the Fermi surfaces of 
the two bands remain isotropic with shifted Fermi wave vectors, but the 
relative phases between the two bands are locked to each other. The relative phase changes by $\pm2 \ell\pi$ 
around the Fermi  surface, as shown in Fig. \ref{fig:FS_beta_1}. 
 From a topological point of view, the $\beta_1$ phase
is a map from the Fermi surface $S^1$ to the $S^1$ manifold of the 
$U(1)$ symmetry group. The non-trivial homotopy group $\pi_1(S^1)$ of this mapping is described by
the Kronecker index, the winding number, which is the angular momentum quantum number $\ell$. 
Under the $\mathbf{T}$ or $\mathbf{C}$ transformation, the Kronecker index 
changes sign. Therefore, the $\beta_1$ phase breaks the $\mathbf{T}$ and $\mathbf{C}$ symmetries 
but preserves their combination. Hence, the $\beta_1$ phase is a type $II$ state
and the topological nature of the Kronecker index guarantees that 
the two degenerate $\mathbf{T}$ symmetry-breaking ground states of the $\beta_1$ phase 
cannot be transformed into each other by any continuous processes.

The $v=0$ line marks the first-order phase boundary between the $\alpha_1$ and $\beta_1$ 
phases. When $u+v<0$, a first-order phase transition to the $\beta_1$ phase occurs 
with decreasing $m$ and this first-order phase boundary meets the second-order one at a 
tricritical point, the black dot in Fig. \ref{fig:phase1}.

\subsection{Hall conductance of the $\beta_1$ phase}
\label{sub:sec:hall_cond}

The $\beta_1$ phase is a type $II$ $\mathbf{T}$-breaking state with a nonzero spontaneous 
Hall conductance $\sigma_{xy}$.
 As shown in Appendix \ref{app:hall_cond}, following the results of Haldane,\cite{Haldane2004}
 the value of the Hall conductance $\sigma_{xy}$ is quantized 
for an insulator but not for a conductor such as the $\beta_1$ phase.
However, we can still relate $\sigma_{xy}$ with the Kronecker index of 
$S^1\rightarrow S^1$.

Applying Eq. \eqref{eq:hall_conductance}, for the $\beta_1$ phase, the integration region 
of the integral is the annular region comprised between the two Fermi surfaces of the two bands.
If the energy difference between the two bands is small, the $z$ component
of $\vec{n}$ defined in Eq. \eqref{eq:vector_n} of Appendix \ref{app:wilson_loop} can be taken 
as a constant. 
Under this approximation, the Hall conductance is
\begin{align}
\sigma_{xy}=&\frac{n_z(1-n_z^2)}{4}\oint_{FS} \frac{\mathbf{dk}}{2\pi} 
\cdot\left(\tilde{n}_x  \nabla_{\mathbf{k}}\tilde{n}_y-\tilde{n}_y  \nabla_{\mathbf{k}}\tilde{n}_x \right),
\label{eq:hall_conductance2}
\end{align}
where $\tilde{n}_x=n_x/\sqrt{1-n_z^2}$ and $\tilde{n}_y=n_y/\sqrt{1-n_z^2}$
can be considered as the $x$ and $y$ components of a two-dimensional unit vector
(by definition, $|n_z|<1$ in $\beta$ phase).
The integral above is taken around the Fermi surface and measures the
Kronecker index of $S^1\rightarrow S^1$, which counts the number of times the 
relative phase winds around the Fermi surface. Notice that the prefactor of the integral,
$n_z(1-n_z^2)/4$ is unquantized and can be changed continuously. Hence, as expected,  $\sigma_{xy}$ 
is not quantized in
this phase. In particular, $\sigma_{xy}$ vanishes if the Fermi surfaces coincide, $n_z=0$. 

\subsection{Effects of  $U(1)\otimes U(1)$ symmetry-breaking interactions}
\label{sub:sec:other_scatterings}

\begin{figure}
\begin{center}
\subfigure[~$\alpha_1$ phase]{\includegraphics[width=0.22\textwidth]{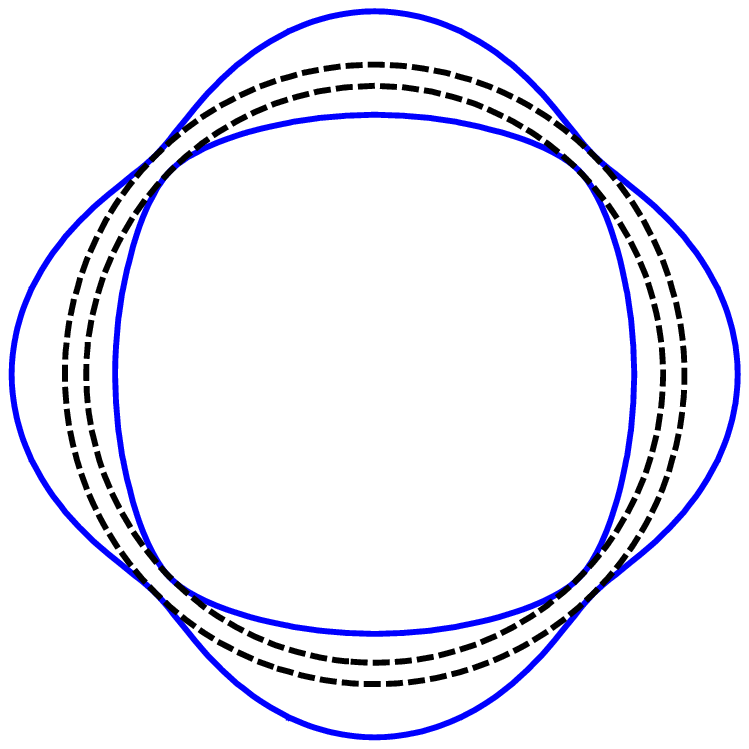}
\label{fig:FS_alpha_1}}
\subfigure[~$\beta_1$ phase]{\includegraphics[width=0.22\textwidth]{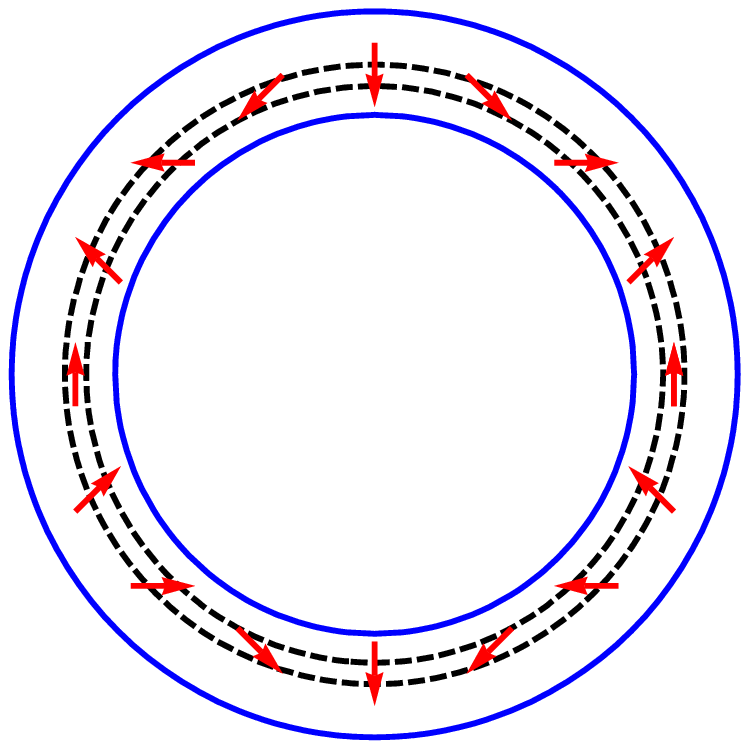}
\label{fig:FS_beta_1}}
\subfigure[~$\alpha_2$ phase]{\includegraphics[width=0.22\textwidth]{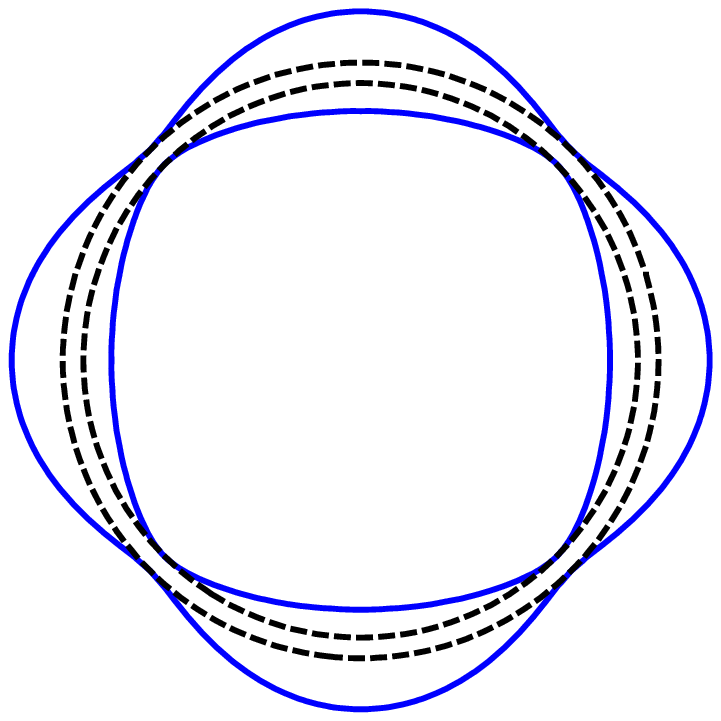}
\label{fig:FS_alpha_2}}
\subfigure[~$\beta_2$ phase]{\includegraphics[width=0.22\textwidth]{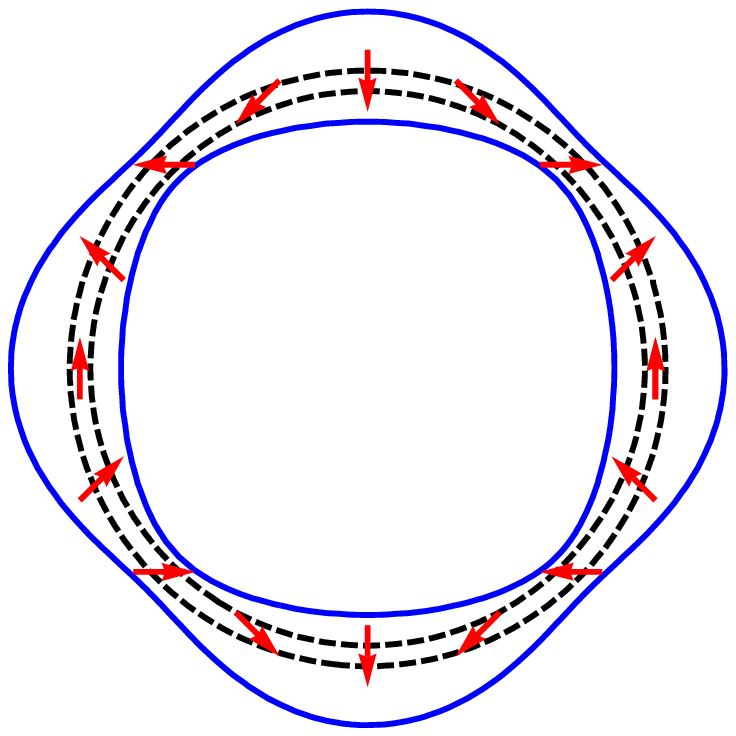}
\label{fig:FS_beta_2}}
\subfigure[~$\alpha_3$ phase]{\includegraphics[width=0.22\textwidth]{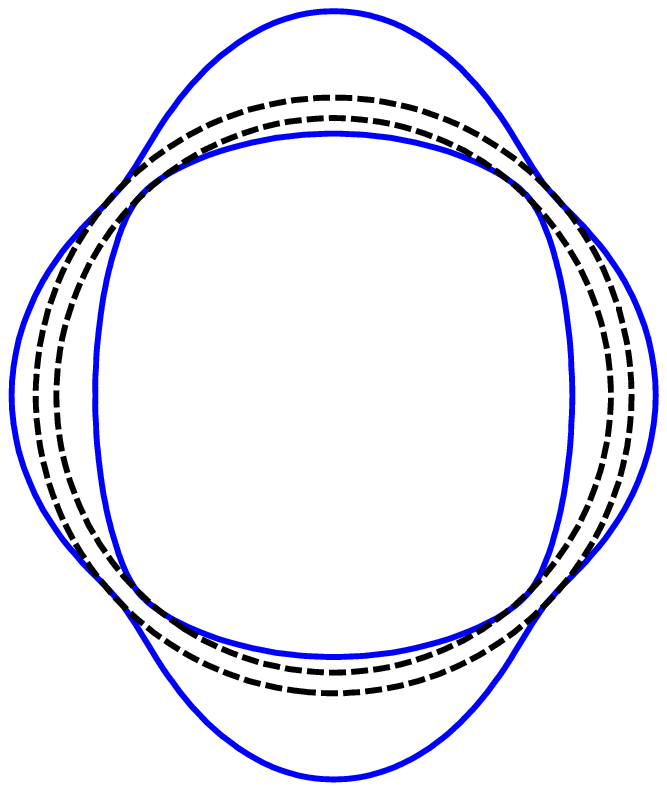}
\label{fig:FS_alpha_3}}
\subfigure[~$\beta_3$ phase]{\includegraphics[width=0.22\textwidth]{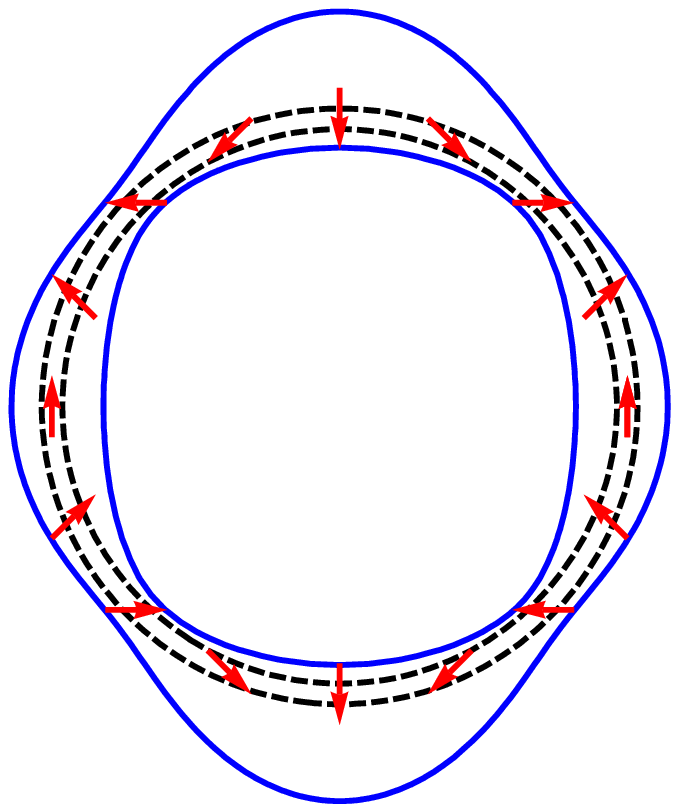}
\label{fig:FS_beta_3}}
\end{center}
\caption{(Color online) The Fermi surface in $\alpha_1$ (a), $\beta_1$ (b), $\alpha_2$ (c), 
$\beta_2$ (d), $\alpha_3$ (e), and $\beta_3$ (f) phases in the $\ell=2$ channel of a 
two-band model. The dashed (solid) lines are the Fermi surfaces of the normal 
(symmetry broken) phases. The small arrows in Fig. (b), (d), and (f) show the relative phases 
of the fermions in the two bands.}
\end{figure}  

We now study the effects of  interactions that were not considered in Sec. \ref{sub:sec:one_scattering}.
The interactions that preserve the particle number in each band, for example, 
$:\psi^\dagger_{1}\psi_{1}::\psi^\dagger_{1}\psi_{1}:$, cannot change qualitatively any conclusions
of Sec. \ref{sub:sec:one_scattering}, since these scattering processes
preserve the relative $U(1)$ symmetry of the two bands.
On the other hand, interactions that change the particle number of each band, such as
$:\psi^\dagger_{1}\psi_{2}::\psi^\dagger_{1}\psi_{2}:$ or 
$:\psi^\dagger_{1}\psi_{1}::\psi^\dagger_{1}\psi_{2}:$,
break the relative $U(1)$ symmetry and potentially can make a difference. However, we can still 
distinguish the two 
different ordered phases, $\alpha$ and $\beta$, depending on whether 
$\vec{\phi}_{\ell 1}\times\vec{\phi}_{\ell 2}$ vanishes or not.

To distinguish from the $\alpha_1$ and $\beta_1$ phases of 
Sec. \ref{sub:sec:one_scattering}, we refer to the $\alpha$ and $\beta$ phases 
for models with $:\psi^\dagger_{1}\psi_{2}::\psi^\dagger_{1}\psi_{2}:$ interactions but not
$:\psi^\dagger_{1}\psi_{1}::\psi^\dagger_{1}\psi_{2}:$ interactions as the $\alpha_2$ 
and $\beta_2$ states. The corresponding phases when both these two kinds of
processes are present will be denoted by  $\alpha_3$ and $\beta_3$, respectively. The $\alpha_3$ 
and $\beta_3$ phases are
the most general ones and do not require fine tuning. 
The main issue we will be interested in is to determine 
if these are genuinely distinct phases, {\it i.e.\/}, if the order parameters have a different 
behavior in all of these cases.

\subsubsection{$\alpha_2$ and $\alpha_3$ phases} 
\label{sec:alpha2alpha3}

Similar to the $\alpha_1$ phases, the $\alpha_2$ and $\alpha_3$ phases preserve 
the $\mathbf{C}$ symmetry. As shown in Figs. \ref{fig:FS_alpha_2} and
\ref{fig:FS_alpha_3}, the Fermi surface of the $\alpha_2$ phase has the same
$2\ell$-fold rotational symmetry as $\alpha_1$, but the $\alpha_3$ phase
has a lower, $\ell$-fold, rotational symmetry. This is because the
$:\psi^\dagger_{1}\psi_{1}::\psi^\dagger_{1}\psi_{2}:$ scattering processes couple the order 
parameters $\phi_{\ell i,x}$ and $\phi_{\ell i,y}$ to the order parameter of the charge 
Pomeranchuk instability in each band
\begin{align}
O_{\ell 1,n}=\avg{\sum_{\mathbf{k}}\psi^\dagger_n(\mathbf{k}) \cos(\ell \theta_k) \psi_{n}(\mathbf{k})},
\\
O_{\ell 2,n}=\avg{\sum_{\mathbf{k}}\psi^\dagger_n(\mathbf{k}) \sin(\ell \theta_k) \psi_{n}(\mathbf{k})},
\end{align}
where $n$ is the band index.
Hence, $O_{\ell 1,n}$ and $O_{\ell 2,n}$ also acquire an expectation value in the $\alpha_3$ phase, 
which reduces the rotational symmetries from $2\ell$-fold down to $\ell$-fold. Therefore,
for $\ell$ odd, the $\alpha_3$ phase is a type $I$ time-reversal symmetry-breaking state.

\subsubsection{ $\beta_2$ and $\beta_3$ phases} 
\label{sec:beta2beta3}

The $\beta_2$ and $\beta_3$ phases break both 
$\mathbf{T}$ and $\mathbf{C}$ symmetries but preserve $\mathbf{C T}$, and hence are type $II$ 
time-reversal symmetry-breaking phases. 
This conclusion becomes obvious if one notices that the $\mathbf{T}$ and $\mathbf{C}$ 
symmetry breakings in the $\beta$ phases is described by a topological index
as shown in Eq. \eqref{eq:hall_conductance2}. Hence, this property should survive even
after adiabatically turning on scattering processes that do not preserve the relative $U(1)$ symmetry.
However, due to the absence of an exact relative $U(1)$ symmetry, the symmetry between $\phi_{\ell i,x}$ 
and $\phi_{\ell i,y}$ is no longer preserved. As a result, these two order parameters cannot become critical
at the same time as one tunes the control parameters. Hence, in order to reach the $\beta_2$ or $\beta_3$ 
phases from the normal Fermi-liquid phase, the system must necessarily either go through a 
sequence of two phase transitions, at which one order parameter at a time will get a non-zero 
expectation value, or there will be a direct first-order transition to a state in which both 
are nonzero. 
In addition, in the $\beta_2$ or $\beta_3$ phases, although the 
two order parameters  $\vec{\phi}_{\ell 1}$ and $\vec{\phi}_{\ell 2}$ are still 
perpendicular to each other, without the protection of the $U(1)$ relative phase symmetry their 
magnitudes are no longer equal. 
Hence, in these phases the Fermi surfaces are no longer isotropic. The $\beta_2$ phase with angular 
momentum channel 
$\ell$ has  a Fermi surface with  $2\ell$-fold rotation symmetry, as shown in Fig. \ref{fig:FS_beta_2}. 
This phase breaks the $\mathbf{T}$ and $\mathbf{C}$ symmetries
as the $\beta_1$ phase does but also has Fermi surface with the same $2\ell$-fold 
rotational symmetry as the $\alpha_1$ and $\alpha_2$ phase. In particularly, in the $\ell=1$ channel, the 
$\beta_2$ phase is a charge nematic state \cite{Kivelson1998,Oganesyan2001} but with broken $\mathbf{T}$ 
and $\mathbf{C}$ symmetry. 
In the $\beta_3$ phase, much as in the case of the $\alpha_3$ state,
the $:\psi^\dagger_1\psi_1::\psi^\dagger_1\psi_2:$ scattering processes reduce the rotational symmetry
to $\ell$-fold, as shown in Fig. \ref{fig:FS_beta_3}. Hence, the $\ell=2$ $\beta_3$ phase 
is a charge nematic state with type $II$ $\mathbf{T}$ symmetry breaking.

\subsubsection{Free energy and phase diagram for the $\alpha_2$ and $\beta_2$ phases}
\label{sec:diagram}

We studied the system without the $U(1)$ relative phase symmetry by adding the following interactions
\begin{align}
&\sum_{\mathbf{q};i=1,2}\frac{g_\ell^{(1)}(q)}{2}\left[\phi_{\ell i,x}(\mathbf{q})\phi_{\ell i,x}(\mathbf{-q})-
\phi_{\ell i,y}(\mathbf{q})\phi_{\ell i,y}(\mathbf{-q})\right]
\nn\\
&+\sum_{\mathbf{q};i=1,2}\frac{g_\ell^{(2)}(q)}{2}\left[\phi_{\ell i,x}(\mathbf{q})\phi_{\ell i,y}(\mathbf{-q})+
\phi_{\ell i,y}(\mathbf{q})\phi_{\ell i,x}(\mathbf{-q})\right].
\end{align}
which correspond to the $:\psi^\dagger_1\psi_2::\psi^\dagger_1\psi_2:$ scattering processes.
In mean-field theory, the Landau free energy becomes
\begin{align}
F=(m+\delta/2)(\phi_{\ell 1,x}^2+\phi_{\ell 2,x}^2)+(m-\delta/2)(\phi_{\ell 1,y}^2+\phi_{\ell 2,y}^2)
\nn\\+u(|\vec{\phi}_{\ell 1}|^2+|\vec{\phi}_{\ell 2}|^2)^2
+4v(|\vec{\phi}_{\ell 1} \times \vec{\phi}_{\ell 2}|)^2\nn\\
+\textrm{higher order terms},
\label{eq:anisotropic-F}
\end{align}
with $u$ and $v$ the same as in Eqs. \eqref{eq:mean_field_u} and \eqref{eq:mean_field_v}
and
\begin{align}
m=&-\frac{N(0)}{4}+\Delta^2 \frac{N(0)}{96}
\left[3\left(\frac{N^{\prime}(0)}{{N(0)}}\right)^2-\frac{N^{\prime\prime}(0)}{{N(0)}}\right]
\nn\\
&-\left(\frac{1}{4 \left(f_\ell(0)+|g_\ell(0)|\right)}+\frac{1}{4 \left(f_\ell(0)-|g_\ell(0)|\right)}\right)
\\
\delta=&-\frac{1}{4 \left(f_\ell(0)+|g_\ell(0)|\right)}+\frac{1}{4 \left(f_\ell(0)-|g_\ell(0)|\right)}.
\label{eq:anisotropic-parameters}
\end{align}
Here $g_\ell=g_\ell^{(1)}+i g_\ell^{(2)}$. In fact, a complex $g_\ell$ adds terms such as 
$\phi_{\ell i,x}\phi_{\ell i,y}$ to the Landau free energy, but upon a suitable rotation, 
the Landau free energy can be transformed into the form of Eq.\eqref{eq:anisotropic-F}. 
The notation is the same as before and the leading higher order term is assumed to be 
$w(|\vec{\phi}_{\ell 1}|^2+|\vec{\phi}_{\ell 2}|^2)^3$ with $w>0$ for simplicity.

\begin{figure}
\begin{center}
\includegraphics[width=0.4\textwidth]{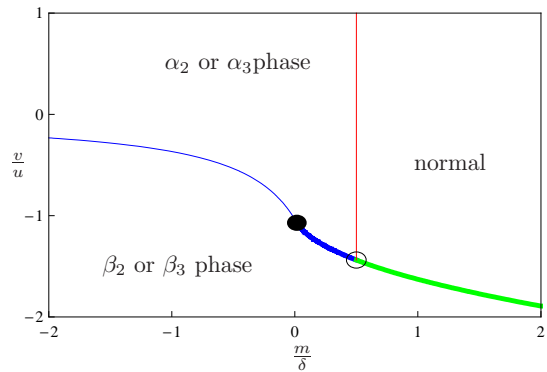}
\end{center}
\caption{(Color online) The phase diagram with broken relative $U(1)$ symmetry.
The notation is the same as in Fig.\ref{fig:phase1}.
The circle represents a
bicritical point where a second-order phase boundary and two first order boundaries meet.}
\label{fig:phase2}
\end{figure} 

The phase diagram is shown in Fig. \ref{fig:phase2}.
As predicted by the general symmetry arguments, the $\alpha_2$ phase can be reached through 
a second-order phase transition at 
$m=\delta/2$. To reach the $\beta_2$ phase one must either go through two transitions 
(using the $\alpha_2$ phase as an intermediate phase) or by a direct first-order transition.
The transition between the $\alpha_2$ and $\beta_2$ phases may be first order or second order depending on details.
In this mean-field theory the first-order and second-order phase boundaries meet at a tricritical point
(the black dot in Fig. \ref{fig:phase2})
located at $m=(2u^2+3 w\delta-2 u\sqrt{u^2+3w\delta})/(12 w)$ and $v=(-u-\sqrt{u^2+3 w \delta})/2$.
The first-order phase boundary between $\alpha_2$ and $\beta_2$ phases meet  the 
second-order phase boundary between $\alpha_2$ and the normal phases, as well as the first-order 
phase boundary between $\beta_2$ and the normal phases, at a bicritical point (the circle in Fig. \ref{fig:phase2}).

For most general interactions, the $\alpha_3$ and $\beta_3$ phases can also occur.
The $\alpha_3$ and $\beta_3$ phases have a similar phase diagram as the
$\alpha_2$ and $\beta_2$ (Fig. \ref{fig:phase2}) and do not exhibit any essentially new phenomena. 
Hence, we will not present here the mean-field study for the $\alpha_3$ and $\beta_3$ phases, 
which has a similar structure to what we have already described in this section.

\section{lattice models}
\label{sec:lattice_model}

Most of the conclusions we reached in the continuum models of the previous section can be 
generalized to the case of lattice models in the absence of degeneracy point. One principal 
difference in the case of lattice models is that the continuum rotational symmetry is 
broken down to a discrete point-group symmetry. In particular this gaps out the corresponding 
Goldstone modes. Another one is that in lattice models the quantum phase transitions to nematic 
states (and their generalizations) often (although this is not necessarily always 
the case as there are a few  known counterexamples) also involve a topological Lifshitz 
transition (from closed to open Fermi surfaces) leading to first-order quantum phase transitions.\cite{Khavkine2004,Yamase2005}
 
\begin{figure}
\begin{center}
\subfigure[]{\includegraphics[width=0.22\textwidth]{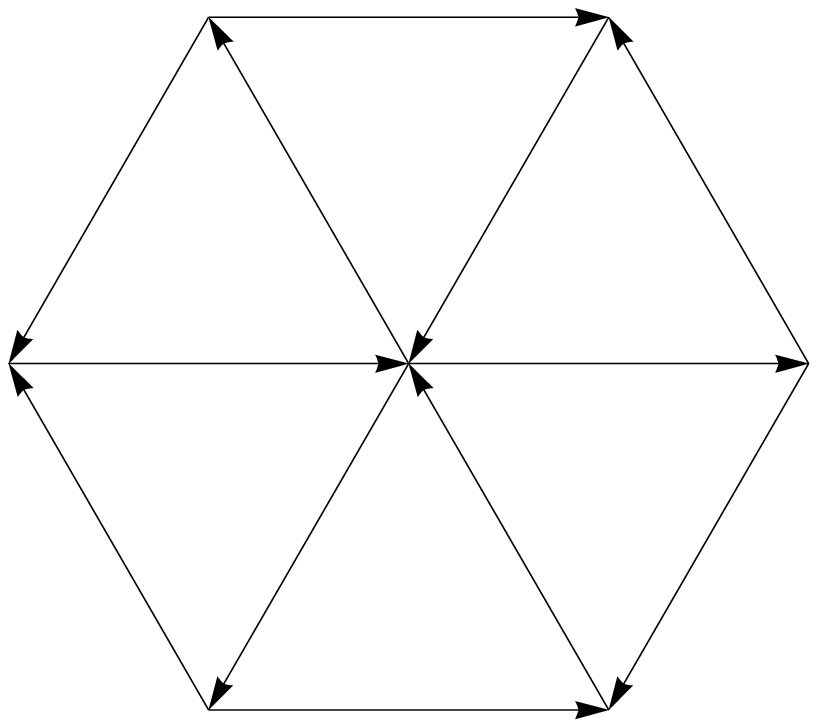}}
\subfigure[]{\includegraphics[width=0.22\textwidth]{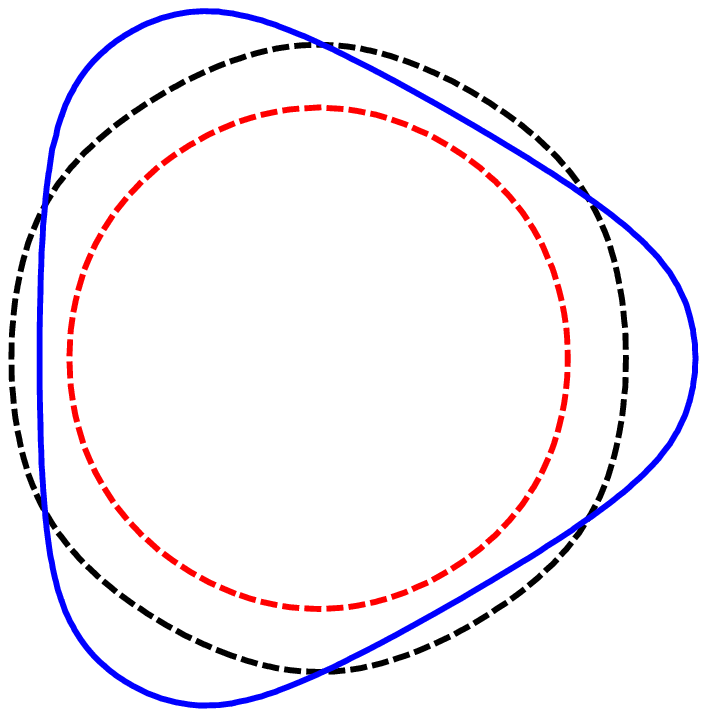}}
\end{center}
\caption{(Color online) Figure (a) is the flux state in a simple triangular lattice
and Fig. (b) shows the Fermi surfaces. The two dashed lines are the Fermi surface
of the states with flux $0$ and $\pm \pi$ (these two states can be transferred into
each other by a gauge transformation) and the solid line is for flux 
$\pm \pi/10$. The rotational symmetry is $3$-fold for flux $\pm \pi/10$ and 
$6$-fold for flux $0$ and $\pi$.
}\label{fig:triangular}
\end{figure}

\begin{figure}[h!]
\begin{center}
\subfigure[]{\includegraphics[width=0.40\textwidth]{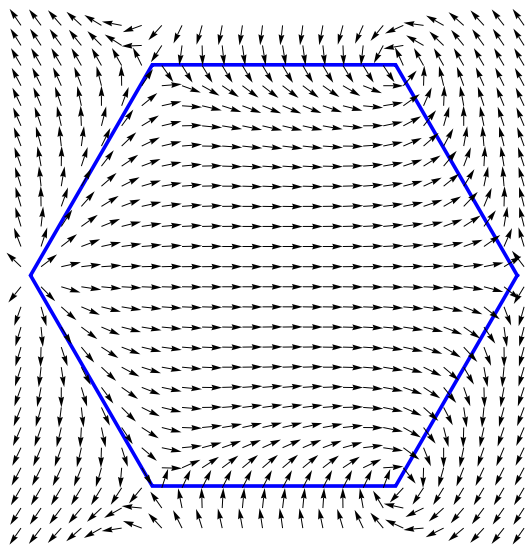}\label{fig:flux_honeycomb}}
\subfigure[]{\includegraphics[width=0.40\textwidth]{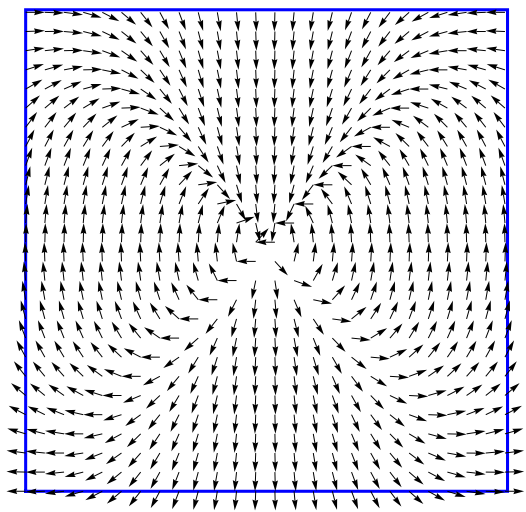}\label{fig:flux_crossed_chain}}
\end{center}
\caption{(Color online) The vector field $(n_x,n_y)$ defined in Eq. \eqref{eq:vector_n} for
the honeycomb lattice, which has $n_z=0$ (a) and the vector field $(n_x,n_z)$ of 
the crossed-chain lattice, which has $n_y=0$ (b). The blue line marks a Brillouin zone. 
For the honeycomb lattice there are two degeneracy points with monopole flux $\pm \pi$ and 
for the crossed-chain lattice, there is one with monopole flux $\pm2\pi$.}
\end{figure}

For one-band models, obviously, the band structure has no degeneracy points. 
Same as in the continuous model, only the type $I$ $\mathbf{T}$ broken-symmetry state can exist. 
As an example, we study a one-band model on 
a simple triangular lattice. This lattice has a $6$-fold rotational symmetry.
With properly chosen interactions, the system can undergo a Pomeranchuk instability 
and form the flux state shown in Fig. \ref{fig:triangular}. 
In this flux state, three currents flow along the three bonds of the simple triangular lattice.
As a result, there is a positive flux in each up-pointing triangle and a negative flux in 
each down-pointing triangle. The net current and the total flux in each unit 
cell are both zero, but the rotational symmetry of the Fermi surface will in 
general be reduced from $6$-fold to $3$-fold symmetry, except when the flux in each 
triangle is $n \pi$ for integer $n$. This is so because a gauge transformation can 
change the flux in a triangle by $2 n \pi$ and similar effects is known to occur for
the square lattice. \cite{Wen1989}  The same analysis with very little modification applies to 
the ``Varma loop model'' $\theta_{II}$ (Ref. \cite{Varma2006}) and to a model on the square lattice with effective 
diagonal hopping terms 
and the same pattern of time-reversal symmetry breaking recently discussed in Ref. \onlinecite{Berg2008}
which are also type $I$ time-reversal symmetry-breaking states.
When the flux is not 
$n \pi$, $\mathbf{T}=\mathbf{I} \ne \mathbf{E}$, where $\mathbf{E}$ is the identity operator.
As shown in Fig. \ref{fig:triangular}, the Fermi surface of a state with flux
different from $n \pi$ has $3$-fold rotational symmetry, which corresponds
to the $\ell=3$ Pomeranchuk instability, and this state belongs to the type $I$ 
$\mathbf{T}$ symmetry-breaking states.

\begin{figure}[h!]
\begin{center}
\subfigure[]{\includegraphics[width=0.20\textwidth]{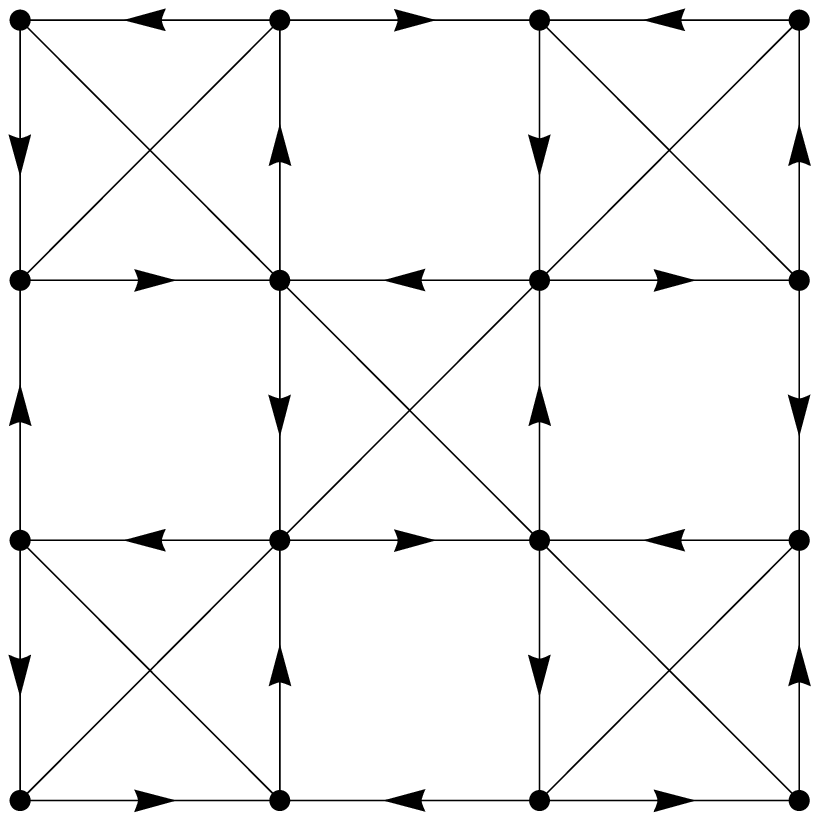}\label{fig:crossed_chain}}
\subfigure[]{\includegraphics[width=0.20\textwidth]{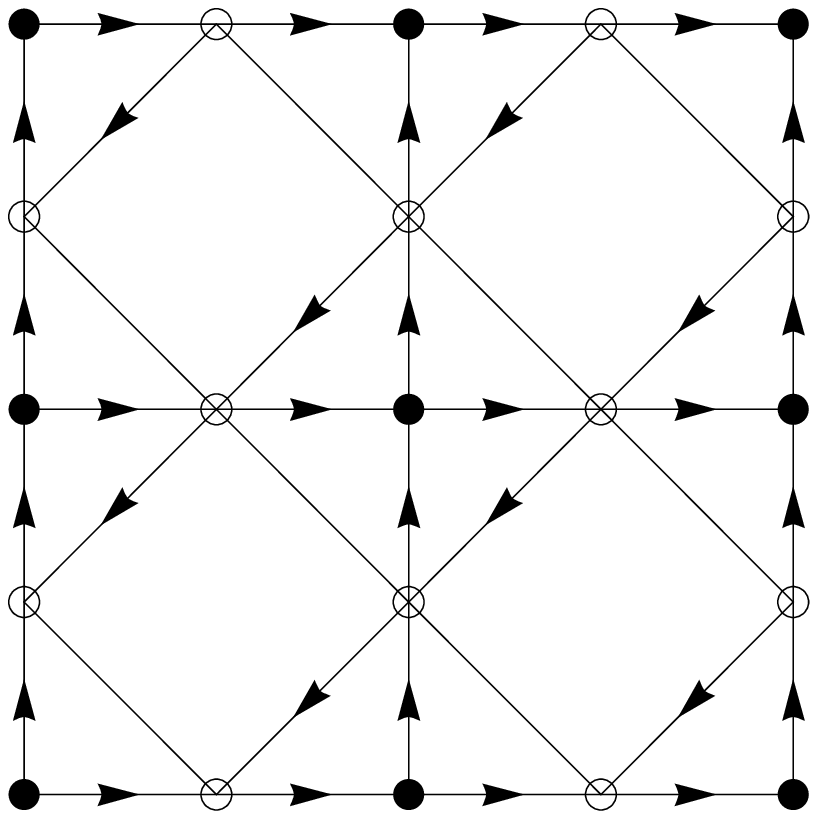}\label{fig:varma_loop}}
\subfigure[]{\includegraphics[width=0.20\textwidth]{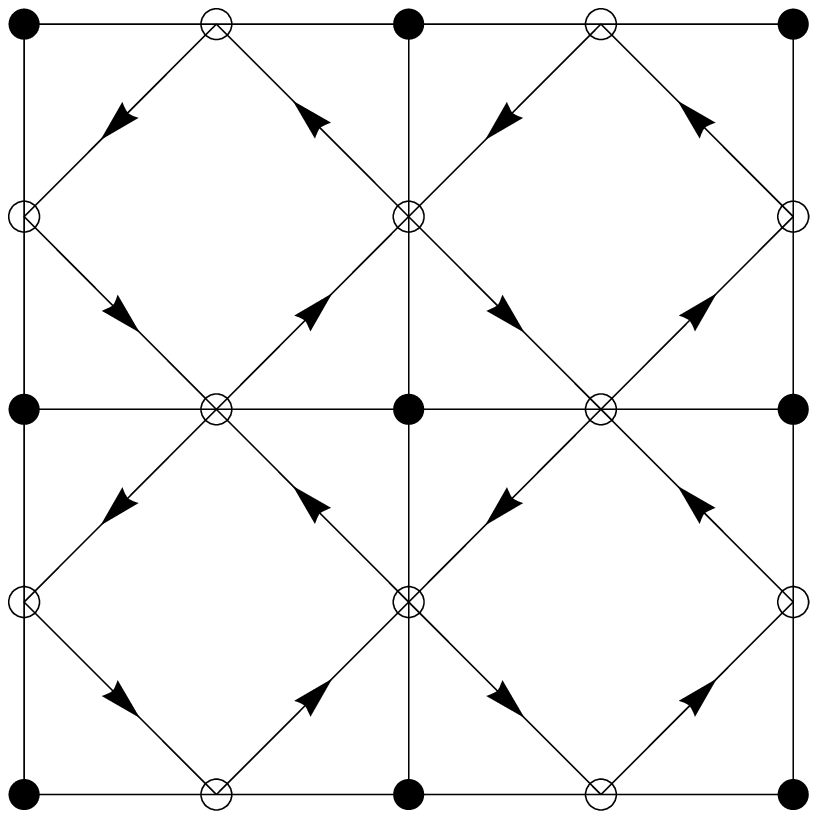}\label{fig:emery_1}}
\subfigure[]{\includegraphics[width=0.20\textwidth]{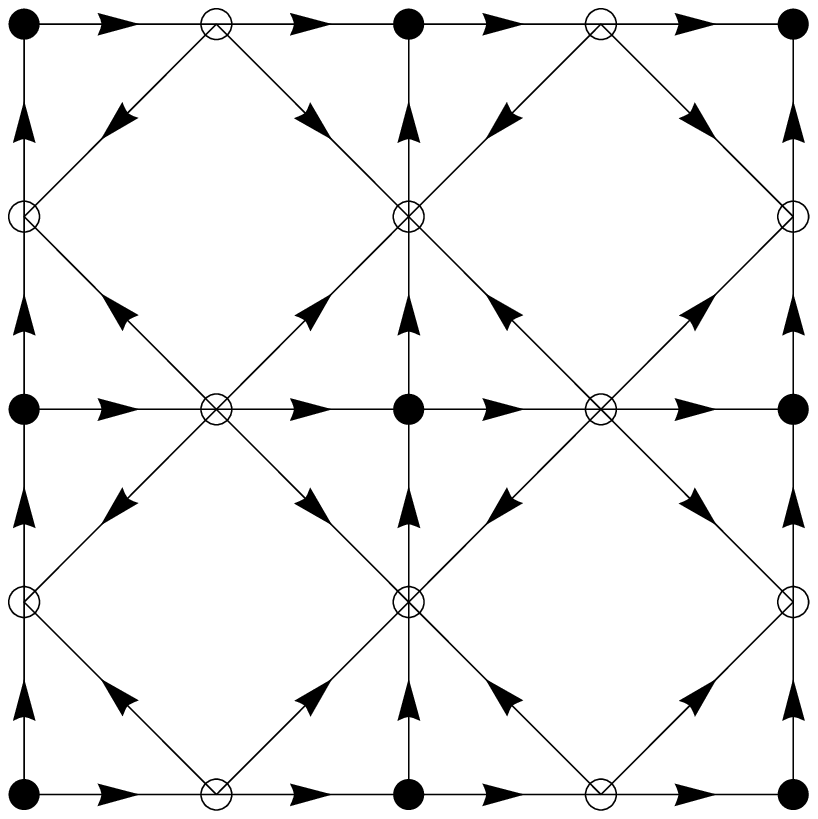}\label{fig:emery_2}}
\end{center}
\caption{The $\mathbf{T}$ symmetry breaking (flux) state in the crossed-chain lattice (a) 
and the Emery lattice (b-d). Fig. (b) is the Varma $\theta_{II}$ loop state. (a) and (c) are type $II$
$\mathbf{T}$-breaking states, (b) is type $I$, and (d) is a type $I$- type $II$ mixed state.}
\end{figure}

In two-band lattice models without degeneracy points, $\alpha$ and $\beta$ phases can
also occur much as in the continuum model discussed before. The simplest example is a bilayer model 
in which two layers of a lattice system are separated by a small distance. If the separation is 
small enough, three dimensional (3D) effects can be ignored. For a bilayer model, 
two bands (bonding and antibonding) can be formed and they usually have no 
degeneracy. In Ref. \cite{Puetter2007},  Puetter, Doh, and Kee studied a system of this type and found a
``hidden nematic phase'' on a bilayer square lattice,
which is one of the example of the $\alpha_1$ phase.

For band structures with degeneracy points, the type $II$ phase
just requires one order parameter quadratic in fermions. A well-known example is
the flux states in the honeycomb lattice. The band structure of the honeycomb
lattice has two degeneracy points (the Dirac nodal points) at the two corners of its first Brillouin zone.
As shown in Appendix \ref{app:wilson_loop} and Fig. \ref{fig:flux_honeycomb}, there 
is a ``monopole flux'' $\pm \pi$ passing through each Dirac point. If the two Dirac points 
get masses with the same sign, they cannot cancel each other; the $\mathbf{C}$ and $\mathbf{T}$ symmetries are broken,
as shown in Fig. 1 of Ref. \cite{Haldane1988}. (A similar effect is found in the mean-field theory of the chiral spin 
liquid \cite{Wen1989,Fradkin1991}.) A recent mean-field study of a system of interacting fermions on the honeycomb 
lattice (at half filling) shows that this state can be stabilized by repulsive next-nearest-neighbor interactions.
\cite{Raghu2008}

Another example is a  fermionic system on the crossed-chain lattice as shown in Fig. \ref{fig:crossed_chain}.
This lattice has two sublattices.\cite{emery-model} The tight-binding model on this lattice has two bands 
and one degeneracy point in each Brillouin zone where the two bands touch. 
This degeneracy point has monopole flux  $\pm 2\pi$
[Fig. \ref{fig:flux_crossed_chain}]. If the four-fold rotation symmetry of the lattice 
is broken explicitly upon the introduction of a different chemical potential for each sublattice,
or spontaneously via a (quantum nematic) symmetry breaking, the degeneracy 
point with flux $\pm 2\pi$ splits into two Dirac points, each with flux $\pm \pi$.
Similar as in the case of  the honeycomb lattice, a Dirac mass term which removes the band crossing breaks
the $\mathbf{T}$ and $\mathbf{C}$ symmetries. 
This state corresponds to the flux state shown in Fig. \ref{fig:crossed_chain} with order parameter
\begin{align}
\sum_\mathbf{k} \sin\frac{k_x}{2}\sin\frac{k_y}{2} \avg{a^\dagger_{\mathbf{k}} b_{\mathbf{k}}-b^\dagger_{\mathbf{k}} a_{\mathbf{k}}},
\end{align}
where $a^\dagger_{\mathbf{k}}$, $a_{\mathbf{k}}$, $b^\dagger_{\mathbf{k}}$, and $b_{\mathbf{k}}$ are the creation and annihilation operators for the two sublattices of the crossed-chain lattice at momentum $\mathbf{k}=(k_x,k_y)$.
It can be shown that this is a type $II$ time-reversal symmetry-breaking state which can be stabilized 
by an arbitrarily weak  nearest-neighbor repulsive interaction for some range of hopping amplitudes and electron densities.\cite{Sun2008} 
Recently, Ran and coworkers \cite{Ran2008}, found similar degeneracy points with $\pm2\pi$ monopole flux in FeAs-based materials with $\mathbf{T}$ symmetry. Hence, similar
$\mathbf{T}$ symmetry-breaking phases may be possible in such systems as well.

The state shown in Fig. \ref{fig:crossed_chain} has a close analogy to the $d$DW state with
$d_{x^2-y^2}+id_{xy}$ symmetry [Fig. 2.(c) in Ref. \cite{Nayak2000} and Fig. 1 in Ref. \cite{Tewari2007}].
Both phases are type II $\mathbf{T}$ symmetry-breaking states. They have similar order parameters but with
very different physical origins. First of all, the lattice structure of our model is a crossed-chain lattice while the $d$DW state is defined on a simple square lattice. Hence, the alternating of the diagonal hopping strength (the crosses) in our model is due to a lattice effect (an explicit symmetry breaking), while in $d$DW state it is due to spontaneous translational-symmetry breaking. Secondly, due to the special lattice structure of the crossed-chain lattice, a non-generic (non-Dirac) band crossing is presented at momentum $(\pi,\pi)$ in the absence of interactions. This band touching leads to an infinitesimal instability to the $\mathbf{T}$ 
symmetry-breaking flux state shown in Fig. \ref{fig:crossed_chain}, where the $d_{x^2-y^2}+id_{xy}$ $d$DW state requires a finite interaction to be reached.

We end the discussion on lattice models with a few remarks on the Emery model, 
which describes the $CuO_2$ plane of cuprates. The Varma loop $\theta_{II}$ state, shown in
Fig. \ref{fig:varma_loop}, is a
type $I$ state that breaks the $\mathbf{I}$ and $\mathbf{T}$ but not the $\mathbf{C}$
and $\mathbf{I T}$.\cite{footnote-varma}
Without breaking translational and 
charge $U(1)$ symmetries, we show in Figs. \ref{fig:emery_1} and \ref{fig:emery_2} two other
nonmagnetic states that 
break the $\mathbf{C}$ symmetry as well as $\mathbf{T}$. The state shown in Fig. 
\ref{fig:emery_1} is the same as the state in Fig. \ref{fig:crossed_chain}, 
if we notice that the oxygens ($p_x$ and $p_y$ orbitals)
in the Emery model form a crossed-chain lattice (rotated by $\pi/4$).
The state in Fig. \ref{fig:emery_2} involves three bands and there is a flux piercing
each small triangle formed by neighboring $d_{x^2-y^2}$, $p_x$, and $p_y$ orbitals.
In each unit cell, three of the triangles have flux $\phi$ but the fourth triangle has flux $-3 \phi$,
with zero total flux in the unit cell. In this state all the
$\mathbf{T}$, $\mathbf{C}$ and $\mathbf{I}$ symmetries are broken (and none of the pairs 
$\mathbf{CT}$, $\mathbf{IT}$, or $\mathbf{CI}$ is preserved) while $\mathbf{TCI}$ remains unbroken.
Hence, this state is a mixture of states of type $I$ and $II$ time-reversal symmetry breaking.

\section{symmetries and finite temperature transitions}
\label{sec:fluctuations}

For a two-band model for a system in the continuum ({\it i.e.\/}, ignoring the explicit breaking of rotational symmetry 
by the underlying lattice), 
the $\alpha_1$ and $\beta_1$ phases break spontaneously two
continuous symmetries: the $SO(2)$ rotational invariance and the relative $U(1)$ phase symmetry. 
Hence, two Goldstone modes are generated in the broken-symmetry phase. At finite
temperatures, the thermal fluctuations of these two Goldstone modes destroy this long-range order. The thermodynamic phase
transition is in the universality class of a system of two $XY$ models which will undergo a Kosterlitz-Thouless type phase
transition.

For the case of the $\alpha_2$ and $\alpha_3$ phases, the relative $U(1)$ phase symmetry of the bands 
is broken explicitly by the effects of the interactions, 
which do not preserve particle number on each band as a good quantum
number. Hence, only one continuous symmetry, $SO(2)$, remains and it is broken spontaneously in this phase at $T=0$. Since
this is a continuous $SO(2)$ symmetry, the finite temperature transition between the normal state and the $\alpha_2$ 
or the $\alpha_3$ phase also belongs to the Kosterlitz-Thouless (KT) universality class, in this case with a single $SO(2)$ order-parameter field.
 
The direct transition from the normal state to the $\beta_2$ or $\beta_3$ states was shown in 
Sec. \ref{sub:sec:other_scatterings} to 
be first order at zero temperature. Therefore one expects this transition to remain first order even at 
finite temperature up
to a critical value where the phase boundary reaches a tricritical (or multicritical) point. 
In general, for a system with full rotational invariance $SO(2)$, the transition from
these phases to the normal state should be in the KT universality class as well.

For lattice systems, rotational symmetry is broken down to a  discrete subgroup, the point-group symmetry of the underlying 
lattice. Hence
the only continuous symmetry left to be spontaneously broken in the $\alpha_1$ and $\beta_1$ phases  is
the relative $U(1)$ phase. Thus, the finite temperature transitions from the $\alpha_1$ and $\beta_1$ phases to the normal
state are also KT transitions. 
For lattice systems in which the $\alpha_2$, $\beta_2$ and $\alpha_3$, $\beta_3$ phases can be realized there are no continuous
symmetries present since the relative $U(1)$ symmetry is broken by the interactions. Hence, in general  the thermodynamic transitions
from the $\beta_2$ and $\beta_3$ phases to the normal state only involve the restoration of time-reversal invariance 
and the discrete point-group 
symmetries broken in these low-temperature phases (up to important caveats discussed below). 

For a lattice system, except for the $\alpha_1$ and $\beta_1$ phases,
whose existence is not generic and requires fine tuning, the $\alpha$ and $\beta$ phases only break
discrete symmetries and  hence do not have Goldstone modes in their excitation spectra. As a result,
the fermionic quasiparticles are the only low-energy excitations
in these broken-symmetry phases. This implies that the Fermi-liquid 
picture should remain  valid in these low-temperature phases even if the fluctuations are stronger 
than what is allowed in a  naive mean-field treatment.

\begin{figure}[h!]
\begin{center}
\subfigure[]{\includegraphics[width=0.22\textwidth]{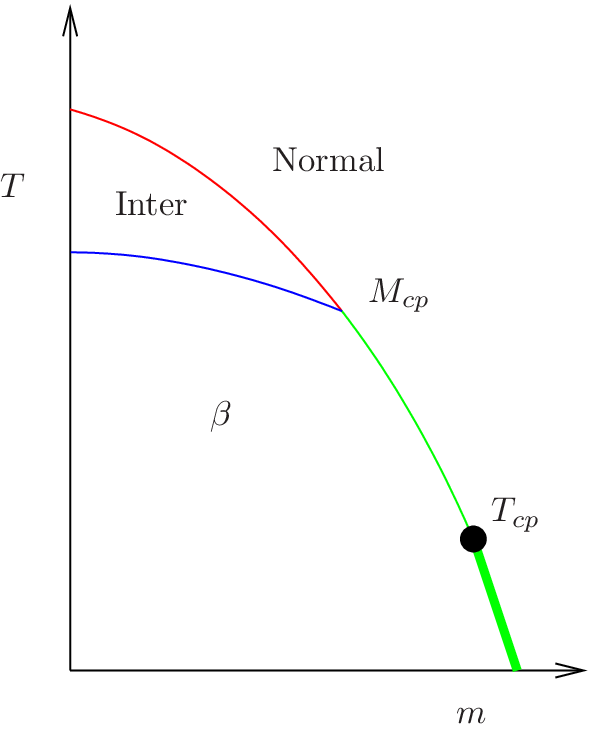}\label{fig:phase-thermal1}}
\subfigure[]{\includegraphics[width=0.22\textwidth]{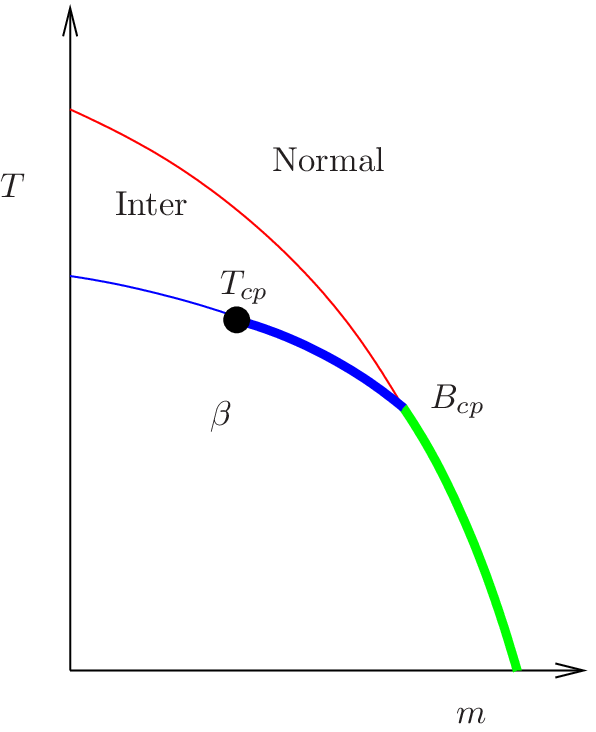}\label{fig:phase-thermal2}}
\subfigure[]{\includegraphics[width=0.22\textwidth]{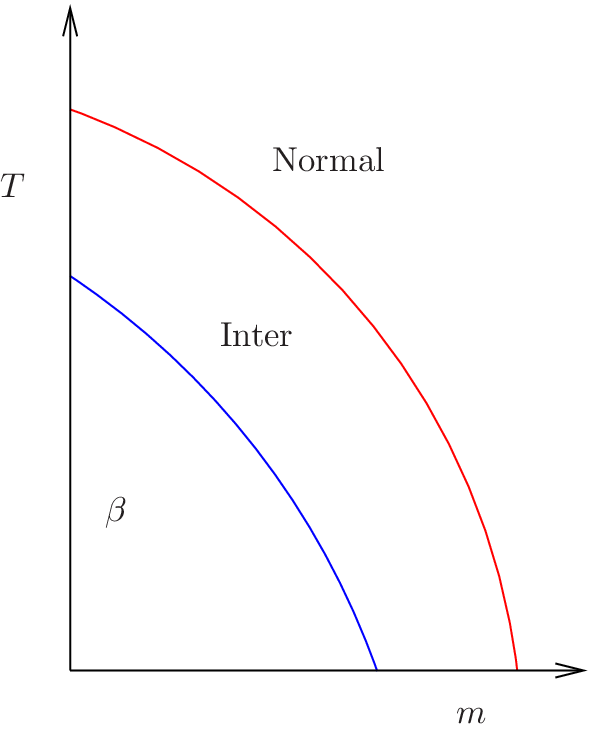}\label{fig:phase-thermal3}}
\end{center}
\caption{(Color online) Different possible schematic phase diagrams for thermal phase transitions
from the $\beta_2$ or $\beta_3$ phases to the normal phase.
Here, ``Inter'' stands for the intermediate phase which may be the $\alpha$ phase or
the isotropic type $II$ time-reversal symmetry breaking phase depending on microscopic details,
although the mean-field approach can only predict the
Fig. \ref{fig:phase-thermal2} and Fig. \ref{fig:phase-thermal3} and requires the intermediate to be the $\alpha$ phase.
The points indicated by $B_{cp}$, $T_{cp}$, and
$M_{cp}$ are the bicritical, tricritical and multicritical points. The thin lines are second
order transitions and the thick lines stand for first order transitions. The convention of the colors
is the same as in Fig. \ref{fig:phase1}. In Fig. \ref{fig:phase-thermal1}, the multicritical point
is a KT phase transition point. The green line between the tricritical and the multicritical points
has continuous varying exponent and all other thin lines in these phase diagrams are Ising
transitions. In Fig. \ref{fig:phase-thermal3}, the phase boundaries (or part of the phase boundary)
may be first order in general.}
\end{figure} 


We will now give a more careful analysis of the nature of the thermal phase transitions for
the $\alpha_2$, $\beta_2$, $\alpha_3$, and $\beta_3$ phases in lattice, which have richer
structures, based on an analysis of the symmetry. 

Let us begin by discussing the d-wave type $\beta_3$ phase in a square lattice. 
In this case, in addition to the $\mathbb{Z}_2$ time-reversal symmetry, this phase also lowers the
invariance under $\pi/2$ spatial rotation to $\pi$ rotation (2D space inversion), which is also 
a $\mathbb{Z}_2$ symmetry breaking. Formally, the $\beta_3$ 
phase has a broken $\mathbb{Z}_2 \otimes \mathbb{Z}_2$ symmetry. 
In the case of the $\beta_2$ phase, the situation is similar except that the
time-reversal even $\mathbb{Z}_2$ symmetry requires a different physical interpretation (see Appendix
\ref{app:Z2}).

From this view point the natural description of the thermal phase transitions out of the $\beta_2$ 
or $\beta_3$ phase should be in the 2D Ashkin-Teller universality class. In what follows we will 
consider the conceptually more straightforward $\beta_3$ phase, but the same analysis also applies 
to the more intricate $\beta_2$ phase.

If the low-temperature
phase is $\beta_3$ the phase diagram can have the general topology of the one shown in Fig. \ref{fig:phase-thermal1}. As
temperature is raised it may undergo a direct phase transition to the normal phase or have two thermal transitions. 
In the latter case, the intermediate phase arises from the restoration of one of the  $\mathbb{Z}_2$ symmetries, either time
reversal or spatial rotations. Thus the intermediate
temperature phase is either a time-reversal even nematic phase (the $\alpha_3$ phase) or a spatially isotropic phase with 
broken time-reversal invariance. In any case, the transition between these 
two states in two space dimensions is continuous and in the 2D Ising universality class.  
One should note that it is possible to conceive the existence of a  spatially isotropic intermediate phase with
broken time-reversal symmetry even at $T=0$. However
a state like that cannot be reached within mean-field theory as it results from the ``quantum melting'' of one of the
$\mathbb{Z}_2$ states of the $\beta$ phase.

These arguments suggest that the finite temperature critical behavior of this system is in general describable 
by a 2D classical Ashkin-Teller model, as  far as
the thermal transitions are concerned. Indeed this would be
the case if the two Ising transitions were to meet at a multicritical point, which would  necessarily be in the 
2D four-state
Potts model universality class, a KT transition. If this scenario is correct, the transition from the $\beta$ 
phase to the 
normal state should have a line of continuous transitions with varying exponents. 
Since the quantum phase transition (at
$T=0$) between the normal state and the $\beta$ phase is first order, this scenario requires the existence of a 
tricritical 
point at some intermediate but low temperature.  A similar phase diagram was found in interacting monomer-dimer
models on a 2D square lattice.\cite{Papanikolaou2007}

An alternative possible scenario is that the two Ising transitions do not meet at a multicritical point. Instead,
that the lower temperature Ising transition becomes first order at some value of the control parameter, 
{\it i.e.\/}, a 2D Ising tricritical point. The general topology of the phase diagram is depicted in 
Fig.\ref{fig:phase-thermal2}. Here too the intermediate phase can either be an $\alpha$ phase, 
in which case time reversal is restored at the lower temperature transition and the intermediate 
phase is nematic, or isotropy is restored first and the intermediate phase breaks time-reversal.

Still a third
possibility arises if at $T=0$ the $\beta$ and normal phases are separated by a region of the $\alpha$ 
phase. In this case  the $T=0$ transition is continuous, and one would generally expect two thermal 
transitions, as shown in Fig.\ref{fig:phase-thermal3}, with an intermediate temperature phase that at 
least at low temperatures must be an $\alpha$ phase. The possibility of another intermediate phase 
discussed above, isotropic and $\mathbf{T}$ breaking, cannot be excluded even in this scenario.

Finally, let us consider briefly the case of $C_{6v}$ lattice symmetry breaking, beyond the $C_{4v}$ symmetry of the
square lattice we have been discussing here, {\it i.e.\/}, simple triangular and honeycomb lattices. We will now have to consider
more general angular momentum channels, with $\ell$ even or odd. Specifically, for the simple triangular and honeycomb lattices the
simplest cases of interest have $\ell=2$ and $\ell=3$ broken by the $C_{6v}$ symmetry. An example of this is the one-band model on a simple triangular lattice discussed above. 

Two-band models on these lattices allow for a richer structure. The simplest case
is a $\beta_3$ state with an $\ell=3$ particle-hole condensate without relative particle-number conservation of any
type. This state breaks spontaneously the $\mathbb{Z}_2$ inversion symmetry and the $\mathbb{Z}_2$ chiral symmetry. 
Hence, the thermal transitions to the normal state of this phase also have a $\mathbb{Z}_2 \otimes \mathbb{Z}_2$, as in the
square lattice case discussed above, with a phase diagram similar to those in
Figs.\ref{fig:phase-thermal1}-\ref{fig:phase-thermal3}. (Just as in the case of the square lattice, this analysis also 
applies to the phase $\beta_2$ with the same caveats on the interpretation of the symmetries.) 

A different symmetry-breaking pattern arises on both the simple triangular and honeycomb lattices if the condensate is in the $\ell=2$ particle-hole channel. In the $\beta_3$ state with $\ell=2$, the $C_6$ symmetry is lowered to $C_2$ (inversion in 2D)
symmetry, where the $C_3$ axis is lifted. Hence, $\mathbb{Z}_3=C_6/C_2$ is the broken rotational symmetry manifold of the order parameter. 
In addition to this $\mathbb{Z}_3$, this phase also breaks the $\mathbb{Z}_2$ time-reversal symmetry, which here is 
equivalent to chiral symmetry. Hence, it breaks a $\mathbb{Z}_3 \otimes \mathbb{Z}_2$ symmetry.
The thermal phase diagram for this $\mathbb{Z}_3\otimes\mathbb{Z}_2$ problem is more complex (and old problem)
than the cases we presented above (see
Refs.[\onlinecite{Cardy1980}] and [\onlinecite{Alcaraz1980}] and references therein). The structure of the phase diagram
can be summarized in three cases: a) a direct transition to the normal state, b) an intermediate temperature critical
phase, and c) an intermediate phase with long-range order. In the first case, 
 the transition between the $\beta_3$ phase and the normal phase is a direct first-order transition 
 (similar to that of the closely related six-state Potts model). In the second case, there is 
a finite range of temperatures in which the system is critical, as in the $\mathbb{Z}_6$ model, and has two KT-type
transitions at each end-point. The third case consists of a sequence of partial restorations of the $\mathbb{Z}_3$ and
$\mathbb{Z}_2$ broken symmetries of the $\beta_3$ state through intermediate temperature phases with either 
$\mathbb{Z}_3$ nematic order and no broken time-reversal symmetry or  with broken time-reversal invariance and full isotropy.

\section{Discussion}
\label{sec:discussion}

We studied microscopic fermionic models (generally in metallic phases) with spontaneous breaking 
of time-reversal invariance.
We considered one-band models and two-band models (ignoring spin) with and without separate 
conservation of particle number
in each band. The time-reversal breaking phases can be classified in two classes:
\begin{enumerate}
\item 
Type $I$ phases, which break time reversal ($\mathbf{T}$) and space inversion ($\mathbf{I}$) but do not break 
chirality ($\mathbf{C}$). We found that type $I$ phases occur in both one-band and multiple-band fermionic systems 
in a generalized  nematic ground state with a (particle-hole) condensate in an 
odd angular momentum channel. Examples of type  $I$ phases we discuss are one-band nematics with $\ell=3$
and two-band models in the $\alpha_3$ state, which has a particle-hole condensate with $\ell$ 
odd without independent particle 
number conservation
in each band.
\item 
Type $II$ phases, which break time reversal  ($\mathbf{T}$) and chirality ($\mathbf{C}$) but do not break space inversion
($\mathbf{I}$). Type $II$ phases have a spontaneous (non-quantized) anomalous Hall effect. 
We found that type $II$ phases are not realized in one-band models. In two (and multiple) band 
models they occur in interband particle-hole condensates with {\em even} angular momentum if particle number is not separately 
conserved in each band ($\beta_3$ phases), \cite{footnote-beta3odd} and in any angular momentum channel $\ell \geq 1$ provided 
particle number is either preserved in each band ($\beta_1$ phases) or conserved modulo $2$ ($\beta_2$ phases).
\end{enumerate}

Each class of states has unique experimental signatures that can be detected in linear and 
nonlinear conductivity measurements, and in the optical response with polarized light.

Let us discuss first how these phases can be detected in transport. It is well known that electronic nematic phases induce
an anisotropy in the conductivity tensor.\cite{Eisenstein1999,Ando2002,Mackenzie2007,Kivelson1998,Oganesyan2001,Kivelson2003} 
In a metallic system, an in-plane electric field $\mathbf{E}$ induces a  current $\mathbf{j}$ which can be expanded 
as a power series in the electric field $\mathbf{E}$,
\begin{align}
j^a = \sigma_{ab} E^b+ \sigma_{abc} E^b E^c+\ldots,
\label{eq:expansion}
\end{align}
with $a$, $b$, and $c$ being $x$ or $y$. The first term in Eq.\eqref{eq:expansion} is the linear response, 
and the second term is the leading nonlinear response.  
The different components of the conductivity tensor (and of the nonlinear response) can be arranged to transform properly
under chiral $\mathbf{C}$ and space inversion $\mathbf{I}$ symmetries, and can be used to 
detect these broken-symmetry phases in experiment. Thus, the conductivity tensor is sensitive to both rotations under
$90^\circ$ and chirality, while the third rank nonlinear conductivity tensor is odd under inversion. 
To detect systems with condensates with $\ell >3$ , it is necessary to  consider higher order nonlinear response terms.
Thus, one can construct phenomenological ``order parameters'' using the electrical response tensors, and the simplest ones are
presented in Table \ref{table:conductance}.  

The signatures of type $I$ $\mathbf T$ breaking
phases can be detected through nonlinear optical processes, such as polarized Raman scattering. 
[They can also be checked directly  by angle-resolved photoemission spectroscopy (ARPES) by detecting the  anisotropy of the Fermi 
surface.]
On the other hand, type $II$ phases, such as the $\beta$ phases, can be detected optically through a nonzero Kerr effect 
(in the absence of external magnetic fields). 

\begin{widetext}
\begin{center}
\begin{table}[h]
\begin{tabular}{|c|c|c|c|c|c|c|c|}
\hline
 &$\sigma_{xx}-\sigma_{yy}$ & $\sigma_{xy}$ & $\sigma_{abc}$
\\
\hline
Normal phase & zero & zero & zero
\\
Nematic ($\ell=2$ phases)  &nonzero& zero & zero
\\
Type $I$ ($\ell=3$ phases, $\mathbf T$ and $\mathbf{I}$ odd) & zero or nonzero & zero & nonzero
\\
Type $II$ ($\beta$, $\ell=2$ phases, $\mathbf{T}$ and $\mathbf{C}$ odd) & zero or nonzero & nonzero & zero
\\
\hline
\end{tabular}
\caption{Conductance tensors for different phases.}
\label{table:conductance}
\end{table} 
\end{center}
\end{widetext}

The problem of constructing phases of electronic systems with broken time-reversal invariance was presented here mainly from its intrinsic
conceptual interest.  Our interest in this 
problem has been to a large extent motivated by the recent discovery of time-reversal symmetry-breaking 
effects in the ruthenates and in the cuprates. We should stress that although the theory we presented in this paper 
as it stand cannot
describe a strongly correlated system, the patterns of symmetry breaking, as well as their consequences, should be of more general validity.
The strongest experimental evidence available to date of time-reversal symmetry breaking in superconductors 
is in the layered compound {\singleRu}.  Kerr effect rotation experiments\cite{xia-2006} and corner 
junction experiments\cite{kidwingira-2006} strongly suggest that this material may indeed  have a 
$p$-wave superconducting state which breaks spontaneously time-reversal symmetry. 
This is consistent with a theoretical prediction\cite{Rice1995} of a condensate with a $p_x+i p_y$ 
order-parameter symmetry. This evidence is however not fully uncontroversial given the conflicting 
experimental results of 
Ref.[\onlinecite{kirtley-2007}] which, so far, have failed to detect the expected edge currents of 
the $p_x+ip_y$ superconductor. 
Recent high-precision Kerr rotation experiments\cite{Xia2007} have now given evidence of weak but 
detectable time-reversal symmetry 
breaking in the underdoped pseudogap regime of the high $T_c$ compound {\YBCO}. 
Neutron-scattering experiments in 
underdoped {\YBCO} \cite{fauque-2006,mook-2008}, and in HgBa$_2$CuO$_{4+d}$ \cite{greven-2008} 
similarly suggest that 
the breaking of time-reversal invariance may also occur in these materials. These recent discoveries 
in the cuprates and 
in {\singleRu} have renewed interest in the possible mechanisms of time-reversal symmetry breaking in 
strongly correlated systems. However, it is worth to note that, in addition to the noted evidence of time-reversal 
symmetry
breaking, neutron-scattering experiments\cite{Hinkov2007} and transport experiments\cite{Ando2002}  find strong
evidence for nematic charge order in {\YBCO} in the same doping range. It is unreasonable to believe that these two phases
can be unrelated to each other, and perhaps they have a common origin. In this sense, the phases found in this paper may
shed some light on these issues.

One important problem that we have not discussed in this work is the role of disorder in these phases. It is well known that disorder couples
as a {\em random field} to the order parameter of anisotropic nematic-like phases,\cite{Carlson2005,Kivelson2003} destroying the
ordered state and rendering the system glassy. The same applies to other phases we discussed here that break spontaneously either point-group
symmetries and/or inversion. On its own, nonmagnetic disorder cannot couple directly to the chiral symmetry and does not destroy
automatically a type $II$ time-reversal breaking state. However, if the spin degrees of freedom are also included even nonmagnetic disorder can
couple indirectly to time-reversal breaking order parameters through the effects of spin-orbit interactions. In this case the system becomes a
$\mathbf{T}$-breaking glassy state. On the other hand, disorder can induce $\mathbf{T}$ and $\mathbf{C}$ breaking effects in phases such as
type $I$ states by breaking locally translation, inversion, and point-group symmetries of the system. In any case, the time-reversal symmetry-breaking effects induced by disorder should be quite weak.\cite{Xia2007}

Other states that break time-reversal symmetry  to 
varying degrees have been postulated 
in the context of high $T_c$ compounds. These states assume the existence of spontaneously circulating 
currents in the ground state. 
They include the  loop state advocated by Varma, \cite{Varma1997,Varma2006} which breaks time reversal 
but not translation invariance, 
and the $d$-density wave state of Charkravarty, Laughlin, Morr and Nayak
\cite{chakravarty01,Nayak2000}, a $d$DW state that  breaks time reversal and translation invariance 
(by one lattice spacing) 
but it is invariant under the 
simultaneous action of both transformations. In the absence of disorder neither of these states exhibits 
a uniform Kerr effect. Tewari {\it et al}\cite{Tewari2007} proposed a $d$DW state with 
$d_{x^2-y^2}+id_{xy}$ symmetry that breaks translational symmetry and 
has nonzero Kerr effect.

There is a close analogy between the $\beta$ phase and the $p_x+i p_y$ (or $d_{x^2-y^2}+i d_{xy}$)
superconducting state. They both need two real order parameters
that couples the $SO(2)$ rotational symmetry and an internal $U(1)$
symmetry to break the $\mathbf{T}$ and $\mathbf{C}$ symmetries.
The $U(1)$ symmetry in the $p_x+i p_y$  spin triplet condensate in the particle-particle 
channel in superconductors is the charge (``gauge'') $U(1)$ symmetry, 
which is an exact symmetry of the system and cannot be broken explicitly. 
On the other hand, the formally analogous $\beta$ phase is a particle-hole condensate which breaks spontaneously the relative
$U(1)$ phase symmetry of a multiband system. This symmetry in general is not exact but it is asymptotically exact  ``emergent''
symmetry at the Fermi-liquid fixed point, since the symmetry-breaking terms are formally irrelevant  (``dangerous
 irrelevant'') operators which are always present in any real system.  
 The difference, between exact and ``emergent'' symmetries, has a direct consequence on the structure of the phase diagram,
 rendering the quantum phase transition from the normal state to the $\beta$ phase first order or
through an intermediate $\alpha$ phase.

Our $\beta$ phases have a close similarity with the phases (with the same name!) in spin-1/2 fermionic systems with 
anisotropic phases such as those discussed in Refs. [\onlinecite{Wu2004}] and [\onlinecite{Wu2007}]. However, the 
$\beta$ phases of those systems are not type $II$ $\mathbf{T}$ symmetry-breaking states 
for two reasons. First under time reversal all three components of the spin polarization of the quasiparticles changes sign.
In contrast, here we have used the Pauli matrices to act on an internal space unrelated with the electron spin, a
``pseudospin'' representing the multiple electronic bands. In our case {\em only} the complex, $\sigma_y$, component 
is odd under time reversal.
Secondly, the internal symmetry  in our problem is only an approximate $U(1)$ phase symmetry, while in the spin 
problem it is the full $SU(2)$ group (in the absence of the spin-orbit couplings). 
We have shown that for systems with $N$ degenerate bands, the (natural) symmetry is $SU(N)$ 
($N=2$ for this case), and the Berry connection $\mathcal{A}^a_{nn}$, defined in Sec. \ref{sec:gauge}, 
is always trivial and  it can always be eliminated by a gauge transformation (it is a gauge transformation!)  
due to Eq. \eqref{eq:constrain}. 
Hence, the actual  Berry phase [no matter $U(1)$ or $SU(2)$] is zero if the $SU(N)$ symmetry is exact. 
As a result, in the fully symmetric case (as well as in the spin-1/2 model) there are no
type $II$ $\mathbf{T}$ symmetry-breaking states. This can also be seen if we notice that the 
map from the Fermi surface to the internal $U(1)$ group is $S^1\rightarrow S^1$, 
which has a nontrivial homotopy group, $\pi_1(S^1)$, with a non-vanishing Kronecker index.
In contrast, for the case of spin-1/2  systems,
the mapping $S^1\rightarrow SU(2)$ has a trivial homotopy, $\pi_1(SU(2))=0$, and does not have a nontrivial 
topological index. We should emphasize that the electronic quasiparticles of the systems we have considered do carry spin-1/2, 
but these degrees of freedom play no role in the phases we have discussed as they 
are paramagnetic.

\begin{acknowledgments}
We thank Steven Kivelson and Congjun Wu for many insightful discussions. 
This work was supported in part by the National Science Foundation
grant DMR 0758462 at the University of Illinois (EF) and  by the Department of Energy, 
Division of Basic Energy Sciences under Award DE-FG02-07ER46453 through the Frederick Seitz Materials
Research Laboratory at UIUC (KS and EF).
\end{acknowledgments}

\appendix
\section{two-band models}
\label{app:two_band}

By definition, the matrix $\mathcal{A}^a_{nm}$
satisfies
\begin{align}
\nabla_\mathbf{k}^a \mathcal{A}^b_{nn}-\nabla_\mathbf{k}^b \mathcal{A}^a_{nn}=
-i \sum_m (\mathcal{A}^a_{nm}\mathcal{A}^b_{mn}-\mathcal{A}^b_{nm}\mathcal{A}^a_{mn}).
\label{eq:constrain}
\end{align}
This constraint on $\mathcal{A}^a_{nm}$ implies that the off-diagonal terms and the diagonal terms are 
related. It follows that
\begin{align}
\sum_{n}(\nabla_\mathbf{k}^a \mathcal{A}^b_{nn}-\nabla_\mathbf{k}^b \mathcal{A}^a_{nn})=0,
\end{align}
which leads to the conclusion that there is no Berry phase in the overall (charge) $U(1)$ sector.

For a two-band model, we can construct three gauge invariant objects that are sensitive to time-reversal 
symmetry breaking
\begin{align}
\nabla_\mathbf{k}^a \mathcal{A}^b_{11}-\nabla_\mathbf{k}^b \mathcal{A}^a_{11},\\
\nabla_\mathbf{k}^a \mathcal{A}^b_{22}-\nabla_\mathbf{k}^b \mathcal{A}^a_{22},\\
i \mathcal{A}^a_{12} \mathcal{A}^b_{21}-i \mathcal{A}^b_{12} \mathcal{A}^a_{21}.
\end{align}
However, due to Eq. \eqref{eq:constrain}, only one of them is linearly independent.

Hence, in the study of the $\mathbf{T}$ symmetry breaking in any two-band model, 
only the diagonal term $\mathcal{A}^a_{11}$ (or $\mathcal{A}^a_{22}$) is needed
to be considered. The same conclusion is trivially valid for one-band models.

\section{the topological and physical meaning of Wilson loops}
\label{app:wilson_loop}

From a topological point of view, in a two-band model,
a unitary transformation at momentum $\mathbf{k}$ belongs to 
the group $U(2)$. By removing the $U(1)\otimes U(1)$ gauge degrees of freedom,
the physical degrees of freedom are in the manifold $U(2)/(U(1)\otimes U(1))=S^2$.
Hence, for each closed loop ${\Gamma}$ in momentum space, a map
$S^1\rightarrow S^2$ can be defined. Similar to the coherent-state
path integral of spin $1/2$ (see for instance Ref. \cite{Fradkin1991}), this map leads to a 
Wess-Zumino term. We will show here the Berry
phase studied in the main text is half of the  Wess-Zumino
term of this map (mod $2  \pi$).

For a two-band model, the kinetic energy part of the Hamiltonian
is a $2\times2$ Hermitian matrix, which can be expanded in
the basis of the identity matrix and the three Pauli matrices:
\begin{align}
H_K(\mathbf{k})=H_0 I+ H_x \sigma_x+ H_y \sigma_y+ H_y \sigma_y.
\label{eq:two_band_Hamiltonian}
\end{align}
Away from degeneracy points, $\sqrt{H_x^2+H_y^2+H_z^2}$ is nonzero.
Hence, we can define a 3D unit vector 
\begin{align}
\vec{n}=\frac{(H_x,H_y,H_z)}{\sqrt{H_x^2+H_y^2+H_z^2}}.
\label{eq:vector_n}
\end{align}
From a topological point of view, this 3D unit vector field
is a map from the momentum space (with degeneracy points removed) to
$S^2$.

If we define the polar coordinates of the unit vector $\vec{n}$ as $\theta$ and 
$\varphi$, where $\vec{n}=(\sin \theta \cos \varphi,\sin \theta \sin \varphi,\cos \theta)$,
the Hamiltonian can be diagonalized by a unitary transformation,
$\mathbf{U}^\dagger H_K(\mathbf{k})\mathbf{U}$,
where
\begin{align}
\mathbf{U}=e^{i \varphi_1}
\left(
\begin{array}{cc}
\cos \frac{\theta}{2} e^{i \varphi_2} & -\sin \frac{\theta}{2} e^{-i\varphi_2+i \varphi} \\
\sin \frac{\theta}{2} e^{i\varphi_2-i \varphi} & \cos \frac{\theta}{2} e^{-i \varphi_2}
\end{array}
\right).
\end{align}
Here $\varphi_1$ and $\varphi_2$ is any function of the momentum, which reflects
the $U(1)\otimes U(1)$ gauge freedom in a two-band model. Therefore, we get
\begin{align}
\mathcal{A}^a_{11}=\nabla_{\mathbf{k}}^a(\varphi_1+\varphi_2-\varphi)+
\frac{1-\cos \theta}{2} \nabla^a_{\mathbf{k}}\varphi.
\end{align}
The loop integral around contour ${\Gamma}$ is
\begin{align}
\Phi_{\Gamma}=\sum_a\oint_{\Gamma} \mathcal{A}_{11}^a dk^a=2 n \pi +
\oint_{\Gamma} \frac{1-\cos \theta}{2} \nabla_{\mathbf{k}}\varphi \cdot d\mathbf{k}.
\end{align}
where $n$ is an integer measuring the winding number of the angle $\varphi_1+\varphi_2-\varphi$.
By comparison, the Wess-Zumino term of the map from ${\Gamma}$ to $S^2$ is
\begin{align}
\iint_B\!\! \vec{n}\cdot(\partial_{k_x}\vec{n} \times \partial_{k_y}\vec{n})  d^2 k
=\!4 n \pi+\oint_{\Gamma} (1-\cos \theta) \nabla_{\mathbf{k}}\varphi \cdot d\mathbf{k}.
\end{align}
Here $B$ is an arbitrary 2D manifold whose boundary $\partial B$ is ${\Gamma}$ and $n$ is an
integer determined by the choice of $B$. Hence, we conclude that the Berry phase 
$\Phi_{\Gamma}$ is half of the Wess-Zumino term up to $2 n \pi$. 

Under a change in chirality, the Wess-Zumino term changes sign (mod $4\pi$).
Therefore, $W_{\Gamma}$ changes into $W_{\Gamma}^*$ under $\mathbf{C}$, which is the same as
$\mathbf{I T}$. Hence, $W_{\Gamma}$ is invariant under $\mathbf{C I T}$.

\section{Hall conductance}
\label{app:hall_cond}

For the Hamiltonian shown in Eq. \eqref{eq:two_band_Hamiltonian}, the
Hall conductance $\sigma_{xy}$ can be evaluated as in Ref. \cite{Yakovenko1990}, using
the three-component unit vectors $\vec{n}(\mathbf{k})$ defined in Eq. \eqref{eq:vector_n},
\begin{align}
\sigma_{xy}=\iint_{B} \frac{\mathrm{d}^2k}{4\pi} \vec{n}\cdot
\left(\partial_{k_x}\vec{n}\times \partial_{k_y}\vec{n}\right).
\label{eq:hall_conductance}
\end{align}
Here the integration region $B$ is the area in momentum space where one
band is filled and the other is empty.

For insulators, the region $B$ is the whole momentum space 
(the first Brillouin zone for lattice models), which is a compact manifold. 
Hence, $\sigma_{xy}$ is
the Kronecker index of the mapping from a compact manifold to $S^2$, the first Chern number, which is
quantized to be an integer. It measures how many times the compact manifold wraps 
around the sphere $S^2$. For conductors, the boundary of $B$ is the Fermi surface, which
implies that $B$ is not a compact manifold. Hence, for conductors in general $\sigma_{xy}$ is not quantized.\cite{Haldane2004}

Using the conclusion from Appendix \ref{app:wilson_loop}, the Hall conductance
$\sigma_{xy}$ can be related to the loop integral $\Phi_{\Gamma}$ as
\begin{align}
\sigma_{xy}=\frac{\Phi_{\Gamma}}{2\pi},
\end{align}
where the contour ${\Gamma}$ is the boundary of $B$, which is the Fermi surface, and
we choose the gauge where $\mathcal{A}^a_{11}$ is analytic in region $B$ to remove
the $2 n \pi$ uncertainty in $\Phi_{\Gamma}$. 
This result is straightforward if we notice that $\Phi_{\Gamma}$
contains all the information of $\mathbf{T}$ and $\mathbf{C}$ symmetry
breaking in two-band models. Therefore, $\sigma_{xy}$, which 
measures the type $II$ $\mathbf{T}$ symmetry breaking at low 
energies, must be directly related with $\Phi_{\Gamma}$, where ${\Gamma}$
is the Fermi surface, which dominates the low-energy physics.

\section{Symmetry analysis of the $\alpha_2$ and $\beta_2$ phases}
\label{app:Z2}

For a $d$-wave-like ($\ell=2$) $\alpha_2$ phase on a square lattice,
the symmetry breaking is reduced from $C_{4v}\otimes \mathbb{Z}_2$  to $C_{2v} \otimes \mathbb{Z}_2$. 
In the normal phase, in addition to the point-group symmetry of the square lattice $C_{4v}$, there is a
internal $\mathbb{Z}_2$ symmetry corresponding to the relative phase shift by $\pi$
between the two bands of a system, in which the relative particle number is conserved mod $2$,
and the relative $U(1)$ phase symmetry is thus reduced to phase shifts of the
electron wave functions between the two bands by $\pi$,
$\psi_1 \rightarrow \psi_1$ and $\psi_2 \rightarrow -\psi_2$. 
In the $\alpha_2$ phase, the order parameter is blind under the simultaneous action of a $\pi/2$ spatial 
rotation and the relative phase shift by $\pi$. Hence, the resulting symmetry of the ordered phase is
$C_{2v}\otimes \mathbb{Z}_2$. Therefore, the thermal  phase transition from the $\alpha_2$ state to the 
normal state is in the 2D Ising universality class. The same analysis applies to the $\alpha_3$ phase, 
except that the broken $\mathbb{Z}_2$ symmetry are simply $C_{4v}/C_{2v}$, which only involves the spatial
symmetry.
 
For the $\beta_2$ phase, in addition to the $\mathbb{Z}_2$ symmetry breaking just discussed above, an 
additional $\mathbb{Z}_2$ time-reversal symmetry is broken. Hence, the $\beta_2$ phase has a broken $\mathbb{Z}_2\otimes\mathbb{Z}_2$ symmetry.


\begin{thebibliography}{52}
\bibitem[{eme()}]{emery-model}
\bibinfo{note}{An effective model with this geometry can be shown to arise in
  the strong-coupling regime of the Emery model of the
  cuprates.\cite{Kivelson2004,Sun2008}}.
\bibitem[{foo({\natexlab{a}})}]{footnote-varma}
\bibinfo{note}{As noted in Sec. \ref{sec:gauge}, Varma's $\theta_I$ loop
  state breaks time reversal but it does not have Berry phases. Hence this
  $\mathbf{T}$ breaking state is neither type $I$ nor type $II$ and it does not
  have a spontaneous anomalous Hall effect.}
\end{thebibliography}
\end{document}